\documentclass[aps,pra,showpacs,twocolumn,superscriptaddress,groupedaddress,longbibliography]{revtex4-1}

\usepackage{graphicx}
\usepackage{amsmath,amssymb}
\usepackage{color}
\usepackage{comment}

\usepackage{cases}  

\usepackage[english]{babel} 

\begin{document}

\title{
BCS superconducting transitions in lattice fermions
}

\author{Kazuto Noda}
\email{noda.kazuto@lab.ntt.co.jp}
\affiliation{NTT Basic Research Laboratories, NTT Corporation, Atsugi 243-0198, Japan}
\author{Kensuke Inaba}
\affiliation{NTT Basic Research Laboratories, NTT Corporation, Atsugi 243-0198, Japan}
\author{Makoto Yamashita}
\affiliation{NTT Basic Research Laboratories, NTT Corporation, Atsugi 243-0198, Japan}

\begin{abstract}
We develop a general description of the superconductivity of lattice fermions based on the BCS theory.
We propose a modeling of the density of states (DOS) of lattice fermions, where divergent and semi-metallic structures are described by asymptotic expansions around the Fermi energy.
This modeling leads to a unified representation of the transition temperature $T_c$ at half filling, which reproduces asymptotic forms of $T_c$ derived in several lattices, such as the square, honeycomb, and Lieb lattices, for the weakly interacting limit.
The derived asymptotic forms of $T_c$ are categorized into four types, which is attributed to the different responses of the degenerate fermions depending on the DOS structures.
The DOS with a delta-functional singularity induces the highest $T_c$ in the weakly interacting region, where $T_c$ is linearly proportional to the pairing interaction $U$.
Three kinds of universal ratios defined in the BCS theory no longer reduce to constants independent of the system parameters but can be parameterized with a certain variable that characterizes the singular structures of the DOS.
We find universal relationship among thermodynamic quantities that holds for all parameter regions.
Further, we numerically demonstrate that in multi-band systems the correlation effects can induce an effective delta-functional singularity.
This phenomenon generally appears in multi-energy (or multi-gap) systems and may provide a plausible guideline for material designs of high-$T_c$ superconductor.
\end{abstract}

\maketitle

\maketitle

\section{Introduction}

Superconductivity (SC) or superfluidity has been one of the most attractive research topics in interacting fermionic systems such as electrons in condensed matter and $^3$He\ \cite{leggett_quantum_2006,parks_superconductivity:_1969}.
Currently, artificially created systems, such as cold atoms \cite{bloch_many-body_2008,regal_observation_2004}, carbon-based materials \cite{emery_superconductivity_2005,ludbrook_evidence_2015}, and layered or thin-film materials \cite{reyren_superconducting_2007,ge_superconductivity_2015}, are activating this research field.
The mechanism of SC transitions in a wide variety of systems can be simply understood on the basis of BCS theory \cite{bardeen_theory_1957}, where SC is described by condensation of Cooper pairs.
This microscopic theory proves that, for instance, transition temperatures can be determined from a few parameters as $T_c\propto e^{-1/\rho(\varepsilon_F) U}$, where $U$  is the paring interaction and $\rho(\varepsilon_F)$ is the density of states (DOS) $\rho(\omega)$ at the Fermi energy $\varepsilon_F$. 
This universal description holds only for weakly interacting regions but is applicable to both continuum and lattice systems.

In lattice systems, the band structure resulting from the discrete translational symmetry could yield a divergent structure of the DOS: $\rho(\omega)\to \infty$. A typical example is a logarithmic divergence in the square lattice $\rho(\omega)\propto \ln(W/|\omega|)$, where $W$ is the bandwidth. 
This divergence is known as a Van Hove singularity (VHS) \cite{van_hove_occurrence_1953}.
When this logarithmic singularity locates around $\varepsilon_F$, the transition temperature derived by the BCS theory is given by $T_c\propto e^{-\sqrt{W/U}}$ \cite{hirsch_enhanced_1986,noda_magnetism_2015}.
Recent experimental progress offers an opportunity to fabricate systems with another type of singularity.
This is a delta-functional singularity (DFS) with $\rho(\omega)\propto \delta(\omega)$ resulting from the (dispersionless) flat bands, and thus it can also be called the flat-band singularity.
In experiments, Kagome \cite{jo_ultracold_2012} and Lieb \cite{taie_matter-wave_2015} lattices with a flat band have been created in cold atom systems with an optical lattice.
In addition, a theoretical study has pointed out that such a singularity appears at the surface of stacked graphene \cite{kopnin_high-temperature_2011}, at that of the strained topological materials \cite{tang_strain-induced_2014}, and in the strained graphene \cite{uchoa_superconducting_2013}.
The BCS theory suggests that this strong singularity yields a linear dependence of $T_c\propto U$ \cite{noda_magnetism_2015}.
In general, the divergent singularities of the DOS change the $U$-dependence of $T_c$.
In particular, a power-law dependence of $T_c$ on $U$ leads to significantly high $T_c$ by comparing with the exponential dependence for weak interactions.
These facts indicate that the control of the DOS, namely, band engineering, paves the way for exploring weakly interacting high-$T_c$ superconductor materials.

Another kind of important DOS structure in lattice systems is the semi-metallic structure with $\rho(\varepsilon_F)\to0$.
For example, in a honeycomb lattice (e.g., graphene), the DOS shows a linear dependence $\rho(\omega)\propto |\omega|$ resulting from the linear dispersion of the Dirac fermions.
The BCS theory predicts that, because of the vanishing $\rho(\varepsilon_F)$, the SC transitions occur at a finite interaction strength $U_c$, and the transition temperatures obey $T_c\propto (U-U_c)$ \cite{kopnin_bcs_2008,uchoa_nodal_2005}.
Because of the finite $U_c$, the BCS theory will somewhat break down, whereas a similar behavior of $T_c \propto (U-U_c)^\beta$ with $\beta\sim0.8$ has been demonstrated by quantum Monte-Carlo simulations \cite{sorella_absence_2012}. Thus, the BCS theory can capture the essence of the transitions of the semi-metallic DOS at a finite $U_c\not=0$.

For both the divergent singular DOS and vanishing semi-metallic DOS, $\rho(\varepsilon_F)$ no longer parameterizes $T_c$, and the well-known form of $T_c\propto e^{-1/\rho(\varepsilon_F) U}$ is insufficient to comprehensively describe the SC transitions in lattice systems.
On the other hand, such an exponential dependence of the characteristic energy scale is a common feature of the weakly interacting fermions with a normal DOS $\rho(\varepsilon_F) \propto 1/W(\not= 0$ and $\infty)$:
such as, the magnetic transition temperatures in lattice systems \cite{hirsch_two-dimensional_1985} and the Kondo temperatures in the impurity systems show the same behavior \cite{hewson_kondo_1993}. 
We thus need to establish general expressions of $T_c$ (and other related quantities such as SC gap energy) by considering the singular structures of the DOS usually appearing in lattice systems.
This generalized theory will allow us to discuss a variety of phase transitions of lattice fermions in the recently created systems mentioned above and also help us to establish a band-engineering material design in the future.

In this paper, we provide a general description of the BCS formulations in lattice systems.
Our approach leads to a unified analytic expression of $T_c$ that reproduces all of the four forms---$T_c\propto e^{-1/\rho(\varepsilon_F) U}$, $\propto e^{-\sqrt{W/U}}$, $\propto U$, and $\propto (U-U_c)$---derived separately as mentioned above.
Our approach further offers a way to comprehensively understand the transitions in lattice systems.
To derive this form, we propose a simple modeling of the DOS to systematically describe the various kinds of the DOS structures.
The unified expressions of the energy gap, thermodynamic quantities such as specific heat, and the universal ratios are derived in the same way.
We clarify that a singular structure of the DOS definitely changes the functional forms of these quantities, whereas we also find universal relationship that holds regardless of the singular structures.
We also present numerical simulations based on the dynamical mean field theory (DMFT), which deals with the local interaction effects precisely.
We here demonstrate that the DFS will effectively appear in multi-energy systems, e.g., when multi-band systems with different bandwidths ($W_{\rm narrow}<W_{\rm wide}$) satisfy the condition $W_{\rm narrow}<U<W_{\rm wide}$.
Systems with such a possible condition will be candidates for the high-$T_c$ weakly interacting superconductors.

Our paper is organized as follows.
In Sec. \ref{sec_gapeq_and_GSDOS}, we introduce a general modeling of the DOS that describes the singular structures.
In Sec. \ref{sec_Tc}, we derive an analytical form of $T_c$ and detail its asymptotic behavior.
In Sec. \ref{sec_Delta}, we calculate the gap energy around $T_c$ and at $T=0$.
In Sec. \ref{sec_universalratios}, we show generalized definitions of the universal ratios dealing with the singular structures of the DOS and relations among these ratios.
In Sec. \ref{sec_effective_flatbandsingularity}, we show the dynamical mean-field results of a model for multi-energy systems to demonstrate the effective delta-functional (flat-band) singularity.
In Sec. \ref{sec_summary}, we summarize our study.

\section{model and method}
\label{sec_gapeq_and_GSDOS}

In this section, we start by explaining our model Hamiltonian and briefly mention the gap equation approach for investigating the SC transitions. 
After that, we propose a modeling of the DOS to systematically deal with divergent and semi-metallic structures.
This approach offers an opportunity to comprehensively understand the SC transitions of lattice fermions.

\subsection{Hubbard Hamiltonian and Gap Equation}

We investigate the SC transitions of lattice fermions and thus consider the Hubbard Hamiltonian.
This simple Hamiltonian describes the physical properties of various systems, such as electrons in condensed matter and atoms in an optical lattice.
This Hamiltonian can be written as
\begin{equation}
\label{eq_hubbard}
{\cal H}
=
\sum_{{\bf k}\sigma} \varepsilon_{\bf k} n_{{\bf k}\sigma}
- U \sum_{i} n_{i\uparrow}n_{i\downarrow},
\end{equation}
where $n_{i\sigma}$ is the number operator of a fermion with spin $\sigma$ in the real space at the $i$th lattice site and $n_{{\bf k},\sigma}$ is that in the momentum space with the wavevector $k$.
$U$ is the pairing interaction strength with $U>0$, and $\varepsilon_{\bf k}$ represents the band dispersion with bandwidth $W$, which we set as a unit of energy.
For simplicity, we avoid detailed discussions of the system parameters, such as the origin of the pairing interaction, the lattice structure, and the form of $\varepsilon_{\bf k}$.
Instead, we consider various functional forms of the local DOS $\rho(\omega)\equiv \sum_{\bf k} \delta(\omega-\varepsilon_{\bf k})$, and we focus only on the asymptotic behavior of the DOS around the Fermi energy $\rho(\omega)\sim a(W/|\omega-\varepsilon_F|)^p+ {\rm const}.$
This simplification (detailed later) allows us to discuss the SC transitions in the various types of lattice systems in a systematic and comprehensive manner.

The following gap equation based on the BCS theory properly describes the s-wave SC transitions \cite{leggett_quantum_2006}:
\begin{equation}
\label{eq_gap}
\frac{1}{U}=\int d\omega  \frac{\rho(\omega)}{2\sqrt{\omega^2+\Delta^2}}\tanh\left(\frac{\sqrt{\omega^2+\Delta^2}}{2T}\right),
\end{equation}
where $T$ is temperature and $\Delta$ represents the superconducting gap energy.
In the next three sections, we analytically solve this equation and determine the functional forms of transition temperature $T_c$, which can be obtained by substituting $\Delta=0$ into Eq.\ (\ref{eq_gap}).
This gap-equation approach is valid for the weakly interacting limit $U\ll W$, which leads to $\Delta \ll W$ and $T_c \ll W$.
We further assume that the band filling is one half, and the lattice structure is bipartite, which yield $\rho(\omega)=\rho(-\omega)$ and $\varepsilon_F=0$.
These assumptions mean that particle-hole symmetry holds, and thus the results in the present study can be straightforwardly applied to the magnetic transitions in bipartite lattices at half filling.
A more general study, such as consideration of systems with asymmetric DOSs, which will require numerical computations, is beyond our current scope.

\begin{figure}[b]
\includegraphics[clip,width=.6\linewidth]{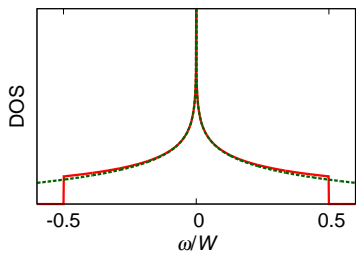}
\caption{(Color online) DOS of the square lattice $\rho^{\text{squ}}(\omega)=\left(2/\pi^2|\omega|\right) K\left(1-(W/2\omega)^2\right) \theta\left(W/2-|\omega|\right)$. The dotted line shows the asymptotic expansion $\rho^{\text{squ}}(\omega)\simeq4\ln(W/2|\omega|)/\pi^2W+8\ln(2)/\pi^2W$.
}
\label{fig_squaredos}
\end{figure}

\subsection{Modeling of the DOS}
We here explain in detail how we deal with the DOS of lattice fermions, which is the key to our approach and leads to unified descriptions of $T_c$, $\Delta$.
As mentioned above, we consider the series expansion of $\rho(\omega)$ around the Fermi energy $\omega \sim \varepsilon_F$.
This is a reasonable extension of the well-known approximation that deals with only the DOS at the Fermi energy without consideration of the functional structures of $\rho(\omega)$ around $\varepsilon_F$.
The main advantage of our approach is that we can comprehensively and systematically deal with various types of DOSs, such as the normal type with a finite $\rho(\varepsilon_F)$, the divergent type $\rho(\omega)\to \infty$, and the semi-metallic (vanishing) type $\rho(\omega)\to 0$.

\subsubsection{Normal DOS}

We first comment on that the simplest modeling provides the well-known solutions of the gap equation for the normal DOS without any singularities around the Fermi energy.
Given the uniform (rectangle-shaped) DOS with step function $\theta(x)$
\begin{equation}
\label{eq_generalizeddos_normal}
\rho^{\text{N}}(\omega)=\frac{1}{W}\theta\left(\frac{W}{2}-|\omega|\right),
\end{equation}
we obtain the conventional exponential form of the transition temperatures: $T_c \propto e^{-1/\rho(\varepsilon_F) U}$, where the DOS at the Fermi energy $\rho(\varepsilon_F)$ is now given by $1/W$.
This conventional modeling successfully captures the essential feature of the SC transition when the DOS is normal metallic with a finite $\rho(\varepsilon_F)$.

\subsubsection{Singular DOS}
We next discuss the modeling of the singular DOS.
In Ref. \cite{van_hove_occurrence_1953}, Van Hove has pointed out that dispersion $\varepsilon_{\bf k}$ is generally flat at the band edges, $\partial\varepsilon_{\bf k}/\partial k_x=0$, and that the DOS of lattice fermions can therefore have singularity $\rho(\omega)\to \infty$. 
A typical example is the logarithmic singularity in the square lattice with dispersion $\varepsilon^{\text{squ}}_{\bf k}= -(W/4)[\cos(k_x)+\cos(k_y)]$.
This logarithmic divergence is called the two-dimensional VHS.
The exact form of the DOS in the square lattice is written as
$\rho^{\text{squ}}(\omega)=(2/\pi^2|\omega|) K \left(1-(W/2\omega)^2\right) \theta\big(W/2-|\omega|\big)$, where $K(x)$ is the complete elliptic integral of the first kind\ \cite{economou_greens_2006}.
An asymptotic expansion of $\rho^{\text{squ}}(\omega)$ around $\omega \sim 0$ shows the logarithmic divergence: $\rho^{\text{squ}}(\omega)\sim (4/\pi^2W) \ln(W/2|\omega|)+8\ln2/\pi^2W+O(\omega^2\ln \omega)$.
Figure \ref{fig_squaredos} shows a comparison between the asymptotic function  of $\rho^{\text{squ}}(\omega)$ and the exact one.
Such an asymptotic expansion adequately captures the structure of the various types of DOSs, which will be generally described as $\rho(\omega)\propto (W/|\omega|)^{p}+{\text{const}.}$
For example, the one-dimensional VHS \cite{van_hove_occurrence_1953} is written as $\rho(\omega)\propto \sqrt{W/|\omega|}$ \footnote{The exact DOS of the one-dimensional system is given by $\rho(\omega)=2/\pi W\sqrt{1-(2\omega/W)^2}$. The one-dimensional VHS appears at the band edges $\pm 2/W$ as $\propto 1/\sqrt{1\mp 2\omega/W}$.}.

We here provide a modeling of the DOS to describe it with a singularity, which is defined as
\begin{eqnarray}
\label{eq_generalizeddos_singular}
\rho^{\text S}(\omega)
&=&
\frac{(1-p)}{pW}
\left[
 \left(\frac{W}{2|\omega|} \right)^{p}-1
\right]
\theta\left(\frac{W}{2}-|\omega|\right),
\end{eqnarray}
where power $p$ of $|\omega|$ characterizes the strength of the singularity.
Singularity strength $p$ should be $p\le 1$ because $\rho^{\text S}(\omega)$ should satisfy the conditions $\int d\omega \rho^{\text S}(\omega)=1$ and $\rho^{\text S}(\omega) \ge 0$ for any $\omega$.

For positive $p$, $\rho^{\text S}(\omega)$ diverges at $\omega=0$ as $(W/|\omega|)^p$: {\it e.g.,} for $p=1/2$,  $\rho^{\text S}(\omega)\propto \sqrt{W/2|\omega|}+{\rm const.}$
It is important to discuss the two special cases of $p=0$ and $1$.
For $p=0$, $\rho^{\text S}(\omega)$ leads to the logarithmic divergence because $\lim_{p\to0}(x^{p}-1)/p= \ln x$, and we obtain $\rho^{\text S}(\omega)=\ln (W/2|\omega|)/W$.
For $p=1$, $\rho^{\text S}(\omega)$ equals $0$ except for $\omega=0$, for which it is indefinite, and we find that $\rho^{\text{S}}(\omega)$ is equivalent to delta function $\delta(\omega)$.
For negative $p$, $\rho^{\text S}(\omega)$ converges as $(|\omega|/W)^{|p|}+{\rm const.}$
For $-2<p<0$, $\rho^{\text S}(\omega)$ still shows a weak singularity: $\rho(\omega \to 0)$ is finite, but shows a kink structure with a discontinuous $\partial\rho(\omega)/\partial \omega$ at $\omega=0$.
This type of weak singularity is also called the VHS.
For example, $p=-1/2$ yields $\rho(\omega) \propto \sqrt{|\omega|/W}+{\rm const.}$ [see Fig.\ \ref{fig_generalizeddos}(b)], and a similar structure of $\rho(\omega)$ can be seen in the cubic lattice.  
This is the three-dimensional VHS.
Note that the negative $p$ is unbounded, and the $p\to -\infty$ limit leads to the uniform rectangular DOS defined in Eq.\ \eqref{eq_generalizeddos_normal} [see Fig.\ \ref{fig_generalizeddos}(e)].

The modeling with the singularity strength $p \in (-\infty, 1]$ makes clear the hierarchical structure in singularity.
The DFS ($p=1$) is the strongest divergence, and the logarithmic singularity ($p=0$) is the weakest.
The power-law singularities $1/|\omega|^p$ are in between.
The negative $p$ shows no divergent singularity, but a weak singularity such as a kink structure still remains.
Note that the kink structure disappears for $p\leq-2$.
The limit of $p\to -\infty$ reproduces the uniform DOS without any functional structures around $\omega=0$.
Furthermore, $d$-dimensional VHS is reproduced with power $p$ satisfying the relation
\begin{equation}
p=1-d/2. \nonumber
\end{equation}
As mentioned above, we find $d=3$ VHS of $\sqrt{|\omega|/W}$ for $p=-1/2$, $d=2$ VHS of $\ln(W/|\omega|)$ for $p=0$, and $d=1$ VHS of $\sqrt{W/|\omega|}$ for $p=1/2$.
The lower the $d$-dimensional singularity is, the stronger the divergence becomes.
The system with the two dimensional VHS ($d=2$) locates a boundary between the systems with divergent and convergent DOSs, and the characteristic logarithmic behavior at $p=0$ reflects this marginal property.
The strongest singularity at $p=1$, the DFS, can be regarded as the zero dimensional VHS ($d=0$).
This type of singularity emerges with or without hopping.
A trivial example is localized fermions without hopping.
On the other hand, certain non-Bravais lattices, such as Lieb and Kagom\'e lattices, have a flat band, which yields a nontrivial DFS with hopping.
We emphasize that the above DOS in Eq. \eqref{eq_generalizeddos_singular} can not reproduce the nontrivial DFS, and thus we further provide a generalized DOS as mentioned below.
We finally note that $d$-dimensional VHSs can be seen in $D$-dimensional lattices, where $d$ does not need to be the same as $D$.
For example, the layered Lieb lattice ($D=3$) contains various VHSs of $d=0, 1, 2,$ and $3$ \cite{noda_magnetism_2015}.

\subsubsection{Generalized singular DOS}

\begin{figure}[ht]
\includegraphics[clip,width=.8\linewidth]{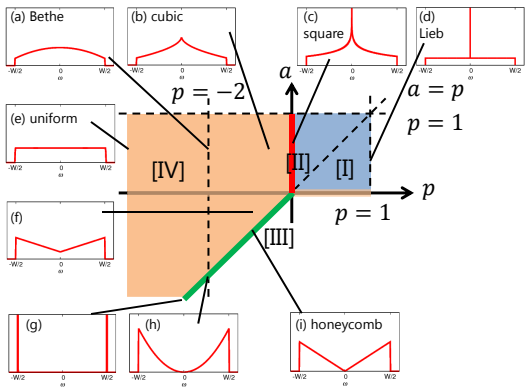}
\caption{(Color online) $p$-$a$ region map for classifing the DOSs of lattice fermions, where $p \in (-\infty, 1]$ and $a \in [\min(p,0) ,1]$.
The four colored regions ${\cal R}_n$ with $n=$ I, II, III, and IV are defined in Table \ref{table_diagram}.
(a)-(i) Typical DOSs with given ($p,a$), some of which asymptotically describe the DOS in the specific lattices: 
(a) ($-2,1/2$) Bethe lattice,
(b) ($-0.5,1/2$) cubic lattice,
(c) ($0,4/\pi^2$) square lattice,
(d) ($1,1/3$) Lieb lattice,
(e) ($-\infty,1/2$) uniform DOS,
(f) ($-1,-1/2$) (typical) concave DOS,
(g) ($-\infty,-\infty$) two delta fucntions \cite{Note2},
(h) ($-2,-2$) depleted DOS,
(i) ($-1,-1$) honeycomb lattice.
}
\label{fig_generalizeddos}
\end{figure}

We further propose the following generalized modeling of the DOS:
\begin{eqnarray}
\label{eq_generalizeddos_simple}
\rho(\omega)
&=&
a\rho^{\text S}(\omega)+b\rho^{\text N}(\omega),
\end{eqnarray}
where $\rho^{\text{S(N)}}(\omega)$ is the singular (normal) DOS defined above in Eq.\ (\ref{eq_generalizeddos_singular}) [(\ref{eq_generalizeddos_normal})].
We introduce singularity weight $a$ with the normalized condition: $\int d\omega \rho(\omega)=a+b=1$.
The condition of $\rho(\omega)\ge 0$ for any $\omega$ requires $\min(p,0) \le a \le 1$.
We should note that the negative weight $a (< 0)$ is possible for negative $p$ with a non-divergent DOS.
Negative weight $a$ means that the DOS shows a concave (or pseudogap) structure around the Fermi energy [see Fig.\ \ref{fig_generalizeddos}(f)].

Using this generalized DOS, we can reproduce the asymptotic expansion of the square lattice with $(p,a)=(0, 4/\pi^2)$, which is shown in Fig.\ \ref{fig_generalizeddos}(c). 
Note that the constant term $(1-a)/W\sim 0.59/W$ does not exactly equal the original value ($8\ln2/\pi^2W\sim 0.56/W$).
This difference is in fact negligible, when the normalized condition $\int d\omega \rho(\omega)=1$ is imposed.
For $p=1$, $\rho(\omega)$ leads to $a \delta(\omega)+(b/W)\theta(W/2-|\omega|)$.
This DOS with $a=1/3$ plotted in Fig.\ \ref{fig_generalizeddos}(d) can describe the DOS of the Lieb lattice as discussed in our previous study \cite{noda_magnetism_2015}, where weight $a$ of one-third means that one of three bands is flat and the other two are dispersive.
Thus Eq.\ \eqref{eq_generalizeddos_simple} with $p=1$ can describe a nontrivial DFS.
For $a=1$ (and $b=0$), $\rho(\omega)$ reduces to the DOS of the trivial localized fermions without hopping: $\rho(\omega)=\delta(\omega)$.
In this way, the combination of the two parameters, singularity strength $p$ and singularity weight $a$, is required to reproduce various types of DOSs of lattice fermions.

The concave DOS with negative $a$ can further represent the semi-metallic DOS.
To clearly show this point, we rewrite Eq. (\ref{eq_generalizeddos_simple}) as
\begin{equation}
\label{eq_generalizeddos}
\rho(\omega)
=
\left\{
a\frac{(1-p)}{p W}
 \left(\frac{W}{2|\omega|} \right)^{ p}
+
\rho_{F}
\right\}
\theta\left(\frac{W}{2}-|\omega|\right),
\end{equation}
where $\rho_{F}=(p-a)/pW$ is a constant term at the Fermi energy.
For $p=a$ with $p<0$, the DOS at the Fermi energy vanishes, $\rho_F=0$, and thus the DOS given by $\rho(\omega)=\frac{(1+|p|)}{W} \left(\frac{2|\omega|}{W} \right)^{|p|}$ describes semi-metallic structures [see Figs.\ \ref{fig_generalizeddos}(g-i)]; for instance, the DOS of the Dirac fermions in a honeycomb lattice is reproduced for $p=-1$ as $\rho(\omega)\propto |\omega/W|$, which is shown in Fig.\ \ref{fig_generalizeddos}(i).
The DOS with $\rho_F=0$ has been studied in a previous work \cite{bacsi_mean-field_2010}.
Note that $\rho_{F}$ is an important parameter that determines various thermodynamic quantities, as will be discussed in the next three sections.

Finally, we summarize how the DOSs of lattice fermions are categorized according to two parameters $a$ and $p$.
As mentioned above, $\rho(0)$ (not $\rho_F$) can be classified into the following three types: the divergent singular ($\rho(0)\to\infty$), the vanishing semi-metallic ($\rho(0)\to0$), and the normal metallic ($\rho(0) =$ finite).
Furthermore, by considering the special property at $p=0$, we distinguish the logarithmic divergence from the other divergences.
We thus provide the following classification with four categories of types of DOSs:
(I) {\it divergent type} in $0<p\leq 1$ and $0<a\leq1$,
(II) {\it logarithmic divergent type} in $p=0$ and $0<a\leq1$,
(III) {\it semi-metallic type} in $p<0$ and $p=a$,
(IV) {\it normal type} in ($p<0$ and $p<a\leq 1$) or $a=0$.
Table \ref{table_diagram} summarizes these four categories  and the corresponding regions ${\cal R}_n$ with $n=$ I, II, III, and IV.
As will be discussed in the next sections, this classification with four categories is convenient when we discuss the properties of the transitions.
Figure \ref{fig_generalizeddos} shows a $p$-$a$ region map for describing the above four DOS categories and provides some examples of DOSs for several choices of $(p,a)$.

\begin{center}
\begin{table}[t]
  \begin{center}
    \caption{Classified regions.}
    \begin{tabular}{c|c} \hline
\label{table_diagram}
       Region &  \\
      \hline\hline
       I & divergent DOS with $\rho(0)=\infty$ \\
         & $\mathcal{R}_{\text{I}}:=$\{$a,p|\  0<p\leq 1 \cap 0<a\leq1$\} \\
       \hline
       II & logarithmic-divergent DOS with $\rho(0)=\infty$ \\
          & $\mathcal{R}_{\text{II}}:=$\{$a,p|\  p=0 \cap 0<a\leq1$\} \\
       \hline
       III & semi-metallic DOS with $\rho(0)=0$ \\
          & $\mathcal{R}_{\text{III}}:=$\{$a,p|\  p<0 \cap p=a$\} \\
       \hline
       IV & normal metallic DOS with finite $\rho(0)$ \\
          & $\mathcal{R}_{\text{IV}}:=$\{$a,p|\ (p<0 \cap p< a\leq 1) \cup a=0\}$ \\	
       \hline
       \hline
       total & ${\cal R}_{\text{tot}} = \mathcal{R}_{\text{I}}\cup\mathcal{R}_{\text{II}}\cup\mathcal{R}_{\text{III}}\cup\mathcal{R}_{\text{IV}}$ \\
          & $ {\cal R}_{\rm tot}:=\{a,p|\ p\leq 1\cap \min(p,0)\leq a \leq 1\}$ \\
       \hline
    \end{tabular}
  \end{center}
\end{table}
\end{center}

\section{Transition temperature}
\label{sec_Tc}

In this section, we discuss how the singular DOS structures affect SC transition temperature $T_c$.
We first explain the physical meaning of the gap equation derived with the generalized singular DOS.
The power-law structure of the DOS around the Fermi energy modifies the gap equation, where we find the competition between polynomial and exponential responses of degenerate fermions as detailed below; that is, the fermions with the normal DOS shows an exponential response such as $e^{-1/\rho_F U}$.
Next, we provide a unified representation of $T_c$ for all systems described by the generalized singular DOS, which is applicable to various lattices, such as the square, honeycomb, and Lieb lattices.
This representation is written down with the Lambert W function $\mathcal{W}_n(x)$, which is the inverse function of $xe^x$ and thus contains both polynomial and exponential properties (for details, see Appendix \ref{appendix_Lambert}).
This duality of $\mathcal{W}_n(x)$ enables us to derive a comprehensive description of the lattice fermion's transitions governed by the Fermi distributions.
Finally, we provide the asymptotic forms of $T_c$ for the weakly interacting limit $U\simeq 0$.
Considering the asymptotic properties of the Lambert W function, we can immediately derive the asymptotic forms of $T_c$ in each region $\mathcal{R}_n$ with $n=$ I, II, III and IV.
These results allow us to systematically and comprehensively understand the BCS transitions of lattice fermions.
Consequently, the high-$T_c$ mechanism originating from the singularities can be parameterized as $p$ and $a$, and then the upper limit of the BCS transition temperature is determined as $T_c=aU/4$ at $p=1$.

\subsection{Physical meaning of the gap equation}
\label{subsec_GapEq}

The following gap equation determines transition temperature $T_c$, which is derived from Eq. (\ref{eq_gap}) with $\Delta=0$ (see Appendix \ref{appendix_DerivationTanhIntegral} for derivations):
\begin{equation}
\label{eq_gap_Tc}
\frac{1}{U}=
-a \chi
+a \frac{B_{T_c}}{W} \left(\frac{W}{T_c} \right)^{p}
+\rho_{F} \ln \left(\frac{1}{A_{T_c}}\frac{W}{T_c} \right),
\end{equation}
where $A_{T_c}$ is a constant $\pi e^{-\gamma_E}$ with Euler's constant $\gamma_E(=0.577)$.
Note that this equation is derived by assuming $T_c/W \ll 1$, and thus we only focus on a few of the leading terms of the $\left(\frac{T_c}{W}\right)$-series.
Here we define the two parameters, $\chi$ and $B_{T_c}$, that depend on $p$:
\begin{eqnarray}
\label{eq_Tc_parameter_Chi}
\chi &=& \frac{1-p}{ p^2}\frac{1}{W} \quad\quad\quad (\geq 0) \\
\label{eq_Tc_parameter_Btc}
B_{T_c}
&=&
\frac{
(1- p) (4-2^{1- p})\Gamma(- p)\zeta(- p)
}{p} \quad (>0),
\end{eqnarray}
where $\zeta(s)$ and $\Gamma(s)$ represent the Riemann zeta and gamma function, respectively.

The gap equation (\ref{eq_gap_Tc}) clearly shows the physical origin of each term.
The third term on the r.h.s with a coefficient $\rho_F$ comes from the uniform structure of the DOS in Eq. (\ref{eq_generalizeddos}).
The exponential dependence of $T_c/W \propto e^{-1/\rho_F U}$ results from this logarithmic term.
This exponential behavior is characteristic of the response of the degenerated fermions around the Fermi surface.
In contrast, the first and second terms with weight $a$ originates from the power-law structure in Eq. (\ref{eq_generalizeddos}).
These terms will lead to the power-law dependence of $T_c/W\propto U^{1/p}$, which characterizes the response of the Fermi surface with the singular DOS.
The coexistence of the exponential and polynomial responses will be a universal physics in the lattice fermions.
Consequently, the competition between these two types of the responses determines the nature of the transitions.

The two constant terms $a\chi$ and $1/U$ in Eq. (\ref{eq_gap_Tc}) can be summed up in a physical term $1/\bar{U}$ defined by:
\begin{equation}
\label{eq_renormalizedinteraction}
\bar{U} = \frac{U}{1+a\chi U}.
\end{equation}
This form allows us to recall the (random-phase-approximation-like) renormalization of the interactions due to many-body effects.
For a small $U\ll W$, the effective interaction $\bar{U}$ is reduced to $\simeq U-a\chi U^2$.
Since $\chi$ is positive, the sign of $a$ determines the renormalization effects.
A convex structure ($a>0$) makes the interaction small, whereas a concave (pseudogap) one ($a<0$) enhances the interaction.
The renormalization is caused by the fermions away from the Fermi energy $\epsilon_F$; for instance, for a concave (convex) structure, more (fewer) particles take part the transitions away from $\epsilon_F$ than those at $\epsilon_F$.
We can see that the renormalization effects disappear for the three limits $a=0$, $p\to -\infty$, and $p=1$ because $a\chi$ goes to zero.
The former two provide the uniform DOS and the latter provides the delta-functional DOS.
Furthermore, in semi-metallic region $\mathcal{R}_{\text{III}}$, this renormalization picture well describes the transition at a finite critical interaction, which is detailed in Sec. \ref{subsec_asymptonic}.

\begin{figure}[tb]
\includegraphics[clip,width=0.8\linewidth]{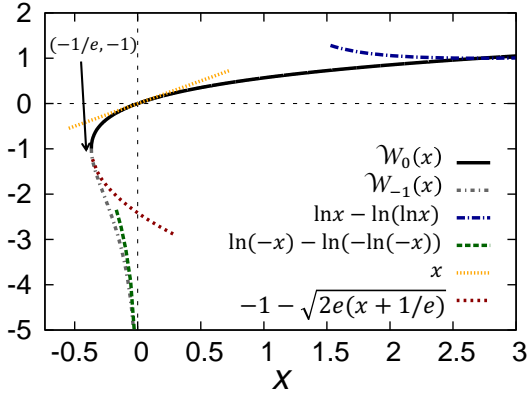}
\caption{(Color online) Lambert W function and its asymptotic expansions. The principal branch $\mathcal{W}_0(x)$ is defined for $-1/e \leq x$ and $-1\leq \mathcal{W}_0(x)$, and the negative branch $\mathcal{W}_{-1}(x)$ for $-1/e \leq x < 0$ and $\mathcal{W}_{-1}(x) \leq -1$. 
Asymptotic expansions are written as $\mathcal{W}_0(x)\sim x \ (x\sim0), \mathcal{W}_0(x)\sim \ln(x)-\ln(\ln(x)) \ (x\to \infty), \mathcal{W}_{-1}(x)\sim -1-\sqrt{2e(x+1/e)} \ (x\sim -1/e)$, and $\mathcal{W}_{-1}(x)\sim \ln(-x)-\ln(-\ln(-x)) \ (x\sim -0)$.
}
\label{fig_lambert}
\end{figure}

\subsection{Unified representation of the transition temperature $T_c$}
\label{subsec_Tc}

We here provide a unified form of transition temperature $T_c$ obtained by solving Eq. (\ref{eq_gap_Tc}):
\begin{eqnarray}
\frac{T_c}{W}
&=&
\label{eq_Tc_2}
\frac{1}{A_{T_c}}
\exp\left(
   -\frac{1}{\rho_{F} \bar{U}}
     +\frac{1}{p}
  	\mathcal{W}_n\left(
	  \frac{p a B_{T_c}A_{T_c}^p}{W\rho_{F}}
          e^{\frac{p}{\rho_{F}\bar{U}}}
           \right)
\right),
\nonumber
\\
\end{eqnarray}
where $\mathcal{W}_n(x)\ (n=0,-1)$ is the Lambert W function defined as the inverse function of $xe^x$.
The coexistence of polynomial and exponential responses mentioned above is properly taken into account by this function.
Figure \ref{fig_lambert} shows the two branches of the Lambert W function defined with a restriction of real-valued $\mathcal{W}_n(x)$ and $x$: the principal branch $\mathcal{W}_0(x)$ is defined for $-1/e \leq x <\infty$ and the negative branch $\mathcal{W}_{-1}(x)$ for $-1/e\leq x <0$.
Note that the Lambert W function is multivalued for a negative $x$. 
The properties of this function are detailed in Appendix \ref{appendix_Lambert}.
We emphasize that a unified expression of $T_c/W$ is derived for all DOS structures described by the generalized singular DOS.
Further, this result enables us to systematically classify the characteristic asymptotic behavior of $T_c$ in each region represented in Table \ref{table_diagram} as discussed below.

We here confirm that the above unified form reproduces the previous results for specific limits.
For $a=0$, where ${\cal W}_0(0)=0$, it reproduces the well-known form $T_c/W=e^{-1/\rho_F U}/A_{T_c}$.
For $p=1$, we obtain $T_c/W=a\big/4b \mathcal{W}_0\left(aA_{T_c}e^{\frac{W}{bU}}/4b\right)$, which is consistent with our previous result \cite{noda_magnetism_2015}.
Note that we here rewrite Eq. (\ref{eq_Tc_2}) to $T_c/W=
\left(\frac{p a B_{T_c}}{W\rho_{F}}\right)^{\frac{1}{p}}
\left[
\mathcal{W}_n\left(
					\frac{p a B_{T_c}A_{T_c}^p}{W\rho_{F}}
          e^{\frac{p}{\rho_{F}\bar{U}}}
\right)
\right]^{-\frac{1}{p}}$
using the definition of the Lambert function $x={\cal W}_n(x)e^{{\cal W}_n(x)}$, and we use $B_{T_c}=1/4$, $\rho_F=b/W$, and $\bar{U}=U$ for $p=1$.
As mentioned below (see Sec. \ref{subsec_asymptonic} and Appendix\ \ref{appendix_asymptoticp}), we can confirm that the $p=0$ limit properly reproduces the logarithmic divergent results \cite{noda_magnetism_2015}.
These $T_c$ results for $p=0$ and $1$ were independently derived in our previous study \cite{noda_magnetism_2015}, where we investigated the layered Lieb lattice that contains both the logarithmic and the delta-functional singularities in the DOS.

\subsection{Asymptotic behaviors of $T_c$} 
\label{subsec_asymptonic}

We here discuss the asymptotic behavior of $T_c$ for the weakly interacting limit $U/W \to 0$.
The unified expression \eqref{eq_Tc_2} elucidates characteristic properties of $T_c$ for each DOS structure: the divergent, log-divergent, semi-metallic, and normal metallic structures.
Considering the limit $U/W\to 0$, we find that the argument of $\mathcal{W}_n(x)$ in Eq.\ \eqref{eq_Tc_2} defined as $X=\frac{apB_{T_c}}{W \rho_F} A_{T_c}^p e^{p/\rho_F \bar{U}}$ goes to different end-points, either $0$, $-1/e$, or $\infty$, depending on $(p,a)$.
We can classify the regions ${\cal R}_{n=\text{I,II,III,IV}}$ defined in Table \ref{table_diagram} to categorize the points to which $X$ goes for $U/W\to0$.
Figure \ref{fig_lambert} shows the asymptotic behavior of ${\cal W}_n(X)$ (see Appendix \ref{appendix_Lambert}), which determines the asymptotic form of $T_c$.
Thus, we can stress that the quantity $\frac{apB_{T_c}}{W \rho_F} A_{T_c}^p e^{p/\rho_F \bar{U}}$ characterizes the response of the degenerated fermions for the weakly interacting limit.
The detailed derivation of the asymptotic forms is shown in Appendix \ref{appendix_asymptoticp}.

We show a summary of asymptotic behavior of $T_c$:
\begin{subnumcases}
{\label{eq_asymptonic} \frac{T_c}{W}=}
 \left(\frac{a B_{T_c}\bar{U}}{W}\right)^{\frac{1}{p}} 
\left(1+ \frac{\bar{U}\rho_F}{p^2}\ln\left(\frac{A_{T_c}^{-p}W}{a B_{T_c}\bar{U}}\right)\right)
\nonumber
\\ 
 \hspace{3.5cm} ({\cal R}_{\text{I}}\cup {\cal R}_{\text{III}})
\label{eq:asymptotic_divergentandsemimetal}
\\
\quad
\simeq
\begin{cases}
\label{eq:asymptotic_divergentandsemimetal2}
\left(aB_{T_c} \frac{U}{W}\right)^{\frac{1}{p}}
& 
 \hspace{0.5cm} (\mathcal{R}_{\text{I}}) \\
 \left(\frac{(U-U_c)W}{|p|B_{T_c}U_c^2}\right)^{\frac{1}{|p|}}
&  
 \hspace{0.5cm} (\mathcal{R}_{\text{III}}) \\
\end{cases} \\
\frac{1}{A_{T_c}}
\exp\left(\frac{b}{a}-\frac{b}{a}\sqrt{1+\frac{2aW}{b^2 U}+\frac{a^2 A_{\text{log}}}{b^2}}\right)
\nonumber
\\
\quad
\simeq
\frac{1}{A_{T_c}}
\exp\left(-\sqrt{\frac{2W}{aU}}\right)
&
\hspace{-2.2cm}($\mathcal{R}_{\text{II}}$)
\label{eq:asymptotic_log}
\\
\frac{1}{A_{T_c}}
\exp\left(
-\frac{1}{\rho_{F} \bar{U}}
\right)
&
\hspace{-2.2cm}($\mathcal{R}_{\text{IV}}$),
\label{eq:asymptotic_normal}
\end{subnumcases}
where $U_c=[p/(p-1)]W$ and $A_{\text{log}}=\gamma_E^2-\pi^2/4+2\ln^2 2+2\gamma_{1}$ with the Stieltjes constant $\gamma_1(=-0.0728)$.
Interestingly, Eq.\ \eqref{eq:asymptotic_divergentandsemimetal} indicates that $T_c$ of the divergent (except for logarithmic divergence) and the semi-metallic DOSs can be described by the same expression, which reflects the fact that the same asymptotic form of ${\cal W}_n(x)$ leads to $T_c$ for both types of the DOSs (see Appendix \ref{appendix_asymptoticp}).
Further considering the sign of $p$, we rewrite it into the different expressions in Eq. (\ref{eq:asymptotic_divergentandsemimetal2}).

The four types of asymptotic forms in Eqs. (\ref{eq:asymptotic_divergentandsemimetal2})-(\ref{eq:asymptotic_normal}) reproduce the previous results \cite{leggett_quantum_2006,hirsch_enhanced_1986,uchoa_nodal_2005,noda_magnetism_2015}.
We should note that our forms are more generalized: $T_c$ can be systematically parameterized by $(a,p)$.
In general, these four forms are not analytical around $U=0$ or $U_c$ (except for some fractional $p$). 
Namely, $T_c$ can not be described by the simple perturbative expansion of $U$ as $\sum C_l U^l$, where $C_l$ is a coefficient with an integer $l$.
This feature is common for both normal and singular DOSs.

In the following, we organize the characteristic of the asymptotic form in each region.
In divergent-structure region $\mathcal{R}_{\text{I}}$, the power-law-$U$ dependence of $T_c$ allows us to expect much higher transition temperatures by comparing them with those of the exponential forms $e^{-1/\rho_F U}$ and also $e^{-\sqrt{2W/aU}}$.
The change in the scalability provides a significantly enlarged $T_c$.
We emphasize that the singularity changes the $U$-dependence of $T_c$ even with a small singularity weight $a$.
The highest $T_c$ of the BCS transitions is given by $T_c\simeq aU/4$, which is induced by the strongest DFS appearing at $p=1$
\footnote{
This functional form has been obtained in an actual system in Ref. \cite{kopnin_high-temperature_2011}.
Kopnin {\it{et al.}} have revealed that a surface SC will be induced by the flattened dispersion at the material surface of stacked graphene, where $T_c$ is linearly proportional to the pairing interaction and the volume of the flattened dispersion corresponding to the singularity weight $a$ in our paper
}. 
We should note that a DFS can be effectively seen in multi-band systems as mentioned in Sec. \ref{sec_effective_flatbandsingularity}.

In semi-metallic structure region $\mathcal{R}_{\text{III}}$, the transition temperature is given by $T_c/W \propto [(U-U_c)W/U_c^2]^{1/|p|}$.
Equation (\ref{eq:asymptotic_divergentandsemimetal}) clearly indicates that the phase transitions occur at finite critical point $U_c\equiv -1/a\chi= [|p|/(|p|+1)]W$, where $p=a<0$.
Note that $U_c$ is defined as the point where the sign of $\bar{U}$ in Eq. (\ref{eq_renormalizedinteraction}) changes (see Appendix \ref{appendix_asymptoticp}).
We should note that, since the interaction is strong at $U\simeq U_c$, correlation effects may spoil our weak-interaction description.
We here compare our result with a previous numerical result for the honeycomb lattice $p=a=-1$, both of which are described as $T_c/W\propto (U-U_c)^{\beta}$.
We obtain $U_c/W=0.5$ and $\beta=1$.
Recent quantum Monte Carlo simulations present $U_c/W=0.645$ and $\beta\sim0.8$ \cite{sorella_absence_2012}, which suggest that our simple mean-field approach provides appropriate estimations.

In logarithmic divergent region ${\cal R}_{\text{II}}$, the transition temperature is given by a specific exponential form $e^{-\sqrt{2W/aU}}$, which yields the higher $T_c$ than the normal exponential form but the lower $T_c$ than the power-law form.
The asymptotic form of $\mathcal{W}_{-1}(x)$ around $x=-1/e$ provides this specific behavior (see Appendix \ref{appendix_asymptoticp}).
It should be emphasized that the logarithmic divergence shows different behavior from the other divergent DOSs, even though the semi-metallic DOSs show the same behavior as the other divergent DOSs.
These features are characteristics of this weakest singularity and convince us that the logarithmic divergence is distinctly categorized from the other divergence.
We further find the characteristic marginal behavior of this singularity as discussed in Sec.\ \ref{sec_universalratios}.

Finally, in normal region ${\cal R}_{\text{IV}}$, the unified form reproduces the conventional form of $T_c\propto e^{-1/\rho_F \bar{U}}$.
Note that, because of the power-law-structure of the DOS around $\epsilon_F$, the interaction is renormalized as $\bar{U}\simeq U-a\chi U^2$.
As mentioned below Eq.\ (\ref{eq_renormalizedinteraction}), the sign of singularity weight $a$ determines whether the interaction is enhanced or suppressed.

\section{Gap energy}
\label{sec_Delta}

This section is devoted to the discussion of the gap energy. 
We first derive $\Delta$ at $T=0$, which shows a similar representaion of $T_c$ in Eq.\ \eqref{eq_Tc_2}.
We next demonstrate that the gap energy around $T_c$ obeys $\Delta(T)\propto \sqrt{(T_c-T)/T_c}$, which indicates that the SC transition is of the second order even with any DOS signularities.
Furthermore, we discuss the behavior of the order parameter $\Phi \equiv \langle c^{\dag}_{i\uparrow} c^{\dag}_{i\downarrow}\rangle$ related with the gap energy as $\Delta=U\Phi$.
With the DFS at $p=1$, $\Phi$ shows characteristic behavior that is relevant for the high-$T_c$ mechanism caused by the DOS singularities.

\subsection{General expression of $\Delta_0$}
\label{subsec_Delta0}

We first calculate the gap energy at zero temperature $\Delta_0$.
We now use the following gap equation obtained at $T=0$, which has the same structure as that of Eq.\ \eqref{eq_gap_Tc}:
\begin{equation}
\label{eq_gap_delta}
\frac{1}{U}=
-a \chi
+a \frac{B_{\Delta_0}}{W} \left(\frac{W}{\Delta_0} \right)^{\bar{p}}
+\rho_{F} \ln \left(\frac{W}{\Delta_0} \right),
\end{equation}
where
\begin{eqnarray}
\label{eq_Delta_parameter_p}
\bar{p}
&=&
\begin{cases}
p 
& (-2< p \leq 1) 
\\
-2
& ( p \leq -2) 
\\
\end{cases}
,
\\
\label{eq_Delta_parameter_BDelta0}
B_{\Delta_0}
&=&
\begin{cases}
 \frac{ \Gamma \left(\frac{3- p }{2}\right) \Gamma \left(\frac{ p }{2}\right)}
   { p 2^{ p} \sqrt{\pi}} 
\quad (>0)
& (-2< p \leq 1) \\
-\frac{a-1}{a}
& ( p =-2) \\
\frac{( p+2-3a)}{a( p+2)}
& ( p<-2)
\end{cases}
.
\end{eqnarray}
We find that the leading term in Eq.\ \eqref{eq_gap_delta} changes at $p=-2$ (see Appendix\ \ref{appendix_int} for details), and as a result, the parameters $\bar{p}$ and $B_{\Delta_0}$ show complex $p$-dependences.
Interestingly, an anomaly at $p=-2$ results from the fact that $\rho(0)$ becomes smooth at $p=-2$; namely, at $\omega=0$, $\rho(\omega)$ has a kink or divergent structure for $p>-2$, whereas it is continuous and smooth for $p\leq -2$.
Such an anomaly in the gap equation cannot be found in that for $T_c$ in Eq.\ \eqref{eq_gap_Tc}.
This difference is attributed to the properties of the steep Fermi surface at $T=0$ and to the smooth one at finite temperatures.

By solving Eq.\ \eqref{eq_gap_delta}, we obtain 
\begin{eqnarray}
\label{eq_Delta}
\frac{\Delta_0}{W}
&=&
\exp\left(
   -\frac{1}{\rho_{F} \bar{U}}
			+\frac{1}{\bar{p}}
  	\mathcal{W}_n\left(
	  \frac{a\bar{p} B_{\Delta_0}}{ W\rho_{F}}
	  e^{\frac{\bar{p}}{\rho_{F}\bar{U}}}
           \right)
\right),
\nonumber
\\
\end{eqnarray}
which is the same as a unified form of $T_c$ when we replace the parameters as follows: $B_{T_c}\to B_{\Delta_0}, A_{T_c}\to 1$, and $p\to \bar{p}$.
Thus, by using this replacement, we obtain asymptotic forms similar to those in Eq.\ \eqref{eq_asymptonic}.
We find that the quantity $\frac{a\bar{p}B_{\Delta_0}}{W \rho_F} e^{\bar{p}/\rho_F \bar{U}}$ shows the same behavior as $\frac{apB_{T_c}}{W \rho_F} e^{p/\rho_F \bar{U}}$, both of which determine the response of the Fermi surface as mentioned above.
In the subset of semi-metallic DOS region ${\cal R}_{\rm III}$, we find a specific behavior of $\Delta_0$: For $a=p\leq-2$, $\Delta_0/W$ is given by $\left[(U-U_c)W/|p| B_{\Delta_0}U_c^2\right]^{1/2}$, where $|p|$ should not be replaced by $|\bar{p}|=2$ because $a=p(\not=\bar{p})$.
A similar anomaly for $a=p\leq-2$ can be found in $\Delta$ around $T_c$ as mentioned below.
It is thus convenient to define regions ${\cal R}^<_{\rm III} :=\{a,p \in {\cal R}_{\rm III}|\ \  p \leq -2\}$ and ${\cal R}^>_{\rm III} :=\{a,p \in {\cal R}_{\rm III}|\ \  p > -2\}$.

\subsection{Gap energy around $T_c$}
\label{subsec_DeltaaroundTc}

We next show that gap energy $\Delta(T)$ around $T_c$ is given for the all of the regions by
$$
\Delta(T)/T_c=R_0 \sqrt{(T_c-T)/T_c}  \quad\quad ({\cal R}_{\rm tot}),
$$
where we introduce the coefficient $R_0$.
This indicates that, even though the singular structures of the DOS change the functional forms of $T_c$, the transitions are of the second order for all of regions $\mathcal{R}_{\text{tot}}$.
(See Appendix \ref{appendix_DeltaNearTc} for details of the calculations shown below.)

With an assumption of $T_c/W \ll 1$, we obtain $R_0$ as
\begin{eqnarray}
\label{eq_DeltaaroundTc_line1}
R_0
&=&
\sqrt{\frac{W\rho_{F}+ a p B_{T_c} (T_c/W)^{- p}}{B_0 W\rho_{F} +a B_1(T_c/W)^{- p}}}
\end{eqnarray}
where
\begin{eqnarray}
B_0&=&\frac{7\zeta(3)}{8 \pi ^2},
\nonumber
\\
B_1&=&\frac{\left(4-2^{- p-1}\right)\left(1- p ^2\right) \sec \left(\frac{\pi   p }{2}\right) \zeta ( p+3)}{ p (2\pi)^{ p+2}},
\nonumber
\end{eqnarray}
which holds for all regions except for ${\cal R}^<_{\text{III}}$.
We can further reduce $R_0$ to
\begin{eqnarray}
R_0&=&
\begin{cases}
\label{eq_DeltaaroundTc}
\sqrt{\frac{p B_{T_c}}{B_1}}
&(\mathcal{R}_{\text{I}} \cup \mathcal{R}_{\text{II}} \cup \mathcal{R}^>_{\text{III}}) \\
\sqrt{\frac{1}{B_0}} (\equiv A_0)
& (\mathcal{R}_{\text{IV}}), \\
\end{cases}
\end{eqnarray}
where the reduced $R_0$ is determined by considering the dominant terms for $T_c/W\ll1$:
for $p \geq 0$ or $\rho_F=0$ (namely, $\in {\cal R}_{\rm I}\cup{\cal R}_{\rm II}\cup{\cal R}_{\rm III}^>$), $(T_c/W)^{-p}$ terms are dominant both in the numerator and denominator, whereas for $a=0$ or $p<0$ (namely, $\in {\cal R}_{\rm IV}$), the constant terms are dominant.

We next discuss $\Delta(T)$ in ${\cal R}^<_{\text{III}}$.
Equation \eqref{eq_DeltaaroundTc} shows that $R_0=\sqrt{p B_{T_c}/B_1}$ vanishes at $p=-2$ and that $R_0=0$ for $p\leq-2$ (namely, $\in {\cal R}^<_{\text{III}}$).
We should note that $\Delta(T)/T_c \propto \sqrt{(T_c-T)/T_c}$ still holds for ${\cal R}^<_{\text{III}}$ with a small but finite coefficint $R_0$:
\begin{eqnarray}
\label{eq_DeltaaroundTc_line3}
R_0
=
\sqrt{\frac{-pB_{T_c}}{B_{\Delta_0}(T_c/W)^{p+2}-B_1}},
\end{eqnarray}
which is further rewritten for $T_c/W \ll 1$ as $\sqrt{-pB_{T_c}/B_{\Delta_0}}(T_c/W)^{-p/2-1}$ for $p<-2$ and $\sqrt{2\pi^2/3\ln(W/T_c)}$ at $p=-2$.
These terms proportional to $(T_c/W)^{-p/2-1}$ or $1/\sqrt{\ln(W/T_c)}$ are very small for $T_c/W \ll 1$.
We neglect such terms in the above expressions in Eq. (\ref{eq_DeltaaroundTc}), and thus $R_0$ is approximately zero at $p=-2$ but is in fact finite for $p\leq -2$.
Note that this anomalous behavior of $R_0$ can be attributed to the smooth and continuous structures of $\rho(\omega)$ for $p\leq -2$.
The semi-metallic DOS with a smoothly decreasing structure means that only a very small number of particles stay around the Fermi energy [see Fig.\ \ref{fig_generalizeddos}(h) and Fig.\ \ref{fig_generalizeddos}(i) for comparison].
We can thus conclude that the depletion of the DOS causes the specific behavior  in ${\cal R}_{\rm III}^<$.

\subsection{Order parameter $\Phi$}
\label{subsec_orderparameter}

The gap energy $\Delta$ is related to the order parameter of the transition $\Phi \equiv \langle c^\dag_{i\uparrow} c^{\dag}_{i\downarrow} \rangle$ as $\Delta=U\Phi$, where $\Phi$ characterizes the number of Cooper pairs.
We find that, with a DFS, at $p=1$, the order parameter $\Phi$ shows the anomalous behavior at zero temperature.
Usually, $\Phi$ goes to zero for the limit of $U/W \to 0$ (or $\to U_c$); {\it e.g.,} in ${\cal R}_{\text{I}}$, $\Phi\propto U^{1/p-1}$ becomes zero, whereas $\Phi$ at $p=1$ stays at a finite value for $U/W\to+0$.
Namely, the number of Cooper pairs suddenly increases from $0$ to $a/2$ with an infinitesimal $U$, which suggests that all of the particles in the flat band form pairs with an infinitesimal $U$.
Note that the same anomalous behavior can be seen in the flat-band magnetism on the Lieb lattice \cite{lieb_two_1989,noda_ferromagnetism_2009}, where $\Phi$ can be mapped onto magnetization $m$.

\section{Universal ratios}
\label{sec_universalratios}

The BCS theory predicts that certain ratios of physical quantities, such as $R_1=2\Delta_0/T_c$, will be constant independent of the system parameters $U$, $\rho_F$, and so on for the wealky interacting limit \cite{bardeen_theory_1957}.
This feature helps us to experimentally demonstrate the BCS transitions \cite{parks_superconductivity:_1969}.
The other examples of the universal ratios are $R_2$ and $R_3$ defined as $R_2=\Delta C(T_c)/\gamma T_c$ and $R_3=H_c(0)^2/\gamma T_c^2$ \cite{bardeen_theory_1957}, where $\Delta C(T)= C_S(T)-C_N(T)$ is the difference in specific heat between the superconducting and the normal conducting states, $\gamma$ is the specific heat coefficient, and $H_c(T)$ is the critical magnetic field.
In the following, we provide general expressions of these universal ratios, including the effects of the singular structures.
For generalization, we rewrite $R_2$ as $\Delta C(T_c)/S(T_c)$ and $R_3$ as $H_c(0)^2/2|F(T_c)|$, where entropy $S(T)$ and free energy $F(T)$ reduce to $\gamma T_c$ and $\gamma T_c^2$ without any singular structures, respectively.
We find that the universal ratios depend only on singularity strength $p$ without weight $a$ dependence.
We further provide a universal relationship $(R_0/R_1)^2=\pi(R_2/R_3)$ that shows no $p$ dependence, where $R_0$ is a coefficient of $\Delta$ near $T_c$ defined in the previous section.
We also discuss how the singular structure affects the thermodynamic quantities $F(T)$, $C(T)$ and $S(T)$.
Appendix \ref{appendix_thermodynamicqunatities} provides a list of thermodynamic quantities and universal ratios of several lattices, such as the square, honeycomb and Lieb lattices, in Table \ref{table_thermodynamics} for convenience.

\subsection{Thermodynamic quantities}
\label{subsec_specificheat_and_entropy}

Before considering the universal ratios, we provide thermodynamic quantities of the normal conductor state: free energy $F_N(T)$, specific heat $C_N(T)$, and entropy $S_N(T)$.
Here, the origin of free energy is set to zero $F_N(0)=0$.
By assuming $T\ll W$, we calculate these quantities as \footnote{
We obtain specific heat $C_N(T)$ from the derivative of internal energy $E(T)[=\int d\omega \omega \rho(\omega) f(\omega)]$, where $f(\omega)$ denotes the Fermi distribution function.
Entropy $S_N(T)$ and free energy $F_N(T)$ are derived from $\int^{T}_{0} dT' \frac{C_N(T')}{T'}$ and $\int^{T}_{0} dT' S_N(T')$, respectively.}
\begin{eqnarray}
\label{eq_specificheat}
C_N(T)
&=&
\frac{2\pi^2}{3}\rho_{F} T
+
a (1-p)B_2
\left(\frac{T}{W}\right)^{1- p},
\\
\label{eq_entropy}
S_N(T)
&=&
\frac{2\pi^2}{3}\rho_{F}T
+
aB_2
\left(\frac{T}{W}\right)^{1- p},
\\
F_N(T)
\label{eq_freeenergy}
&=&
-\frac{\pi^2}{3}\rho_{F}T^2
-
\frac{aB_2}{(2-p)}
\left(\frac{T^{2- p}}{W^{1- p}}\right),
\\
B_2&=&
\frac{\left(2^{2- p}-2\right) \Gamma (3- p )
   \zeta (2- p )}{ p}.
\nonumber
\end{eqnarray}
The first terms with a coefficient $\rho_{F}$ in Eqs. \eqref{eq_specificheat}-\eqref{eq_freeenergy} represent the conventional Fermi liquid behavior, for instance, $C_N(T)\propto T$, which is dominant for $a=0$ or $p<0$ (namely, $\in {\cal R}_{\rm IV}$).
On the other hand, the second terms with $(T/W)^{1- p}$- or $(T/W)^{2- p}$- dependence become dominant in $\mathcal{R}_{\text{I}}\cup\mathcal{R}_{\text{III}}$.
For $p=0$, {\it e.g.}, $C_{N}$ reduces to $\frac{2\pi^2}{3}(1-a) T/W + a\frac{2\pi^2}{3} T\ln(W/T)$ \cite{hirsch_enhanced_1986}, where a $T\ln (1/T)$ term is dominant for $a\not =0 (\in {\cal R}_{\rm II})$ and a linear term is dominant for $a=0 (\in {\cal R}_{\rm IV})$.
These results suggest that singular structures of the DOS induce unconventional $T$-dependence of the thermodynamic quantities.
This unconventional Fermi liquid behavior can be understood as the typical responses of the Fermi surface with the singular DOSs in the same way as above (see Secs.\ \ref{sec_Tc} and \ref{sec_Delta}, the discussion on the gap equations).

For $p=1$ with a DFS, we find characteristic behaviors $C_N(T)=(2\pi^2T/3W)(1-a)$ and $S_N(T)=(2\pi^2/3)(1-a)T/W+ 2a \ln 2$,
which means that the flat-band structure does not take part in the specific heat and that it shows a constant entropy without $T$ dependence.
Large constant entropy of $2a\ln 2$ will be released by the SC transitions at low temperatures; otherwise, the residual entropy remains at zero temperature.
This entropy enhanced by the DOS singularity will be a reason for the high-$T_c$ mechanism.

\subsection{$R_1$: universal ratio}
\label{subsec_universalratio}

We calculate the universal ratio defined as $R_1\ (=2T_c/\Delta_0)$.
This value is constant $A_1$ ($=2\pi e^{-\gamma_E}=3.53$) without any singular structures of the DOS, which is one of the important consequences of the BCS theory. 
This statement should be extended to deal with singular DOSs.
We thus elucidate that the universal ratio $R_1$ can be described by a function of the singular strength $p$ without an $a$ dependence.
This suggests that the structure of $\rho(\omega)$ around the Fermi energy cannot be neglected in discussing the phase transitions even though singularity weight $a$ is small.

Using Eqs.\ (\ref{eq_Tc_2}) and (\ref{eq_Delta}), we obtain
\begin{eqnarray}
\label{eq_R}
R_1&\equiv& 2\Delta_0/T_c \nonumber \\
&=&
{2A_{T_c}}
\exp
\left(
 \frac{1}{\bar{p}}
	\mathcal{W}_n\left(
	  \frac{a \bar{p} B_{\Delta_0}}{W\rho_{F}}
          e^{\frac{\bar{p}}{\rho_{F}\bar{U}}}
     \right)
\right.
\nonumber
\\
&&
\hspace{2cm}
\left.
- \frac{1}{{p}}
\mathcal{W}_n\left(
	  \frac{a p B_{T_c}}{W\rho_{F}}
		A_{T_c}^{p}
          e^{\frac{p}{\rho_{F}\bar{U}}}
     \right)
\right)
\nonumber
\\
&=&
\begin{cases}
2\left(\frac{B_{\Delta_0}}{B_{T_c}}
\right)^{1/ p}
&(\mathcal{R}_{\text{I}} \cup \mathcal{R}_{\text{II}} \cup \mathcal{R}_{\text{III}}^>) \\
0
& (\mathcal{R}_{\text{III}}^<) \\
2A_{T_c}(=A_1)
& (\mathcal{R}_{\text{IV}}). \\
\end{cases}
\end{eqnarray}
For the singular DOS structure regions ($\mathcal{R}_{\text{I}} \cup \mathcal{R}_{\text{II}} \cup \mathcal{R}_{\text{III}}^>$)
$R_1$ can be described by a single representation with a $p$ dependence.
On the other hand, in normal metallic region $\mathcal{R}_{\text{IV}}$, $R_1$ is reduced to the conventional value $A_1(=2A_{T_c})$.
In depleted structure region $\mathcal{R}_{\text{III}}^<$, we obtain $R_1\propto (\frac{T_c}{W})^{-p/2-1} \simeq +0$ with $T_c/W\ll 1$ for $a=p\leq-2$ (see Appendix \ref{appendix_depletionDOSregion}).
This means that $T_c\gg\Delta_0$ in depleted DOS region $\mathcal{R}_{\text{III}}^<$, because the depletion of the DOS is crucial at zero temperture with the steep Fermi surface, while it has a little effect at finite temperatures.

\begin{figure}[tb]
\includegraphics[clip,width=0.8\linewidth]{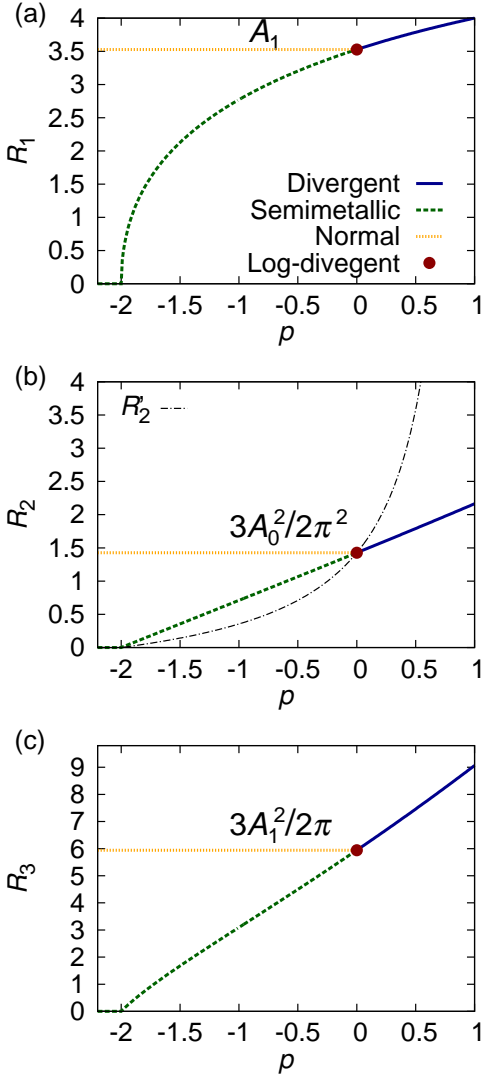}
\caption{(Color online) (a) $R_1$, universal ratio,
(b) $R_2$, specific heat jump at $T=T_c$,
(c) $R_3$, critical magnetic field at $T=0$.  
}
\label{fig_universalratio}
\end{figure}

Figure\ \ref{fig_universalratio}(a) shows the $p$ dependence of $R_1$.
In ${\cal R}_{\text{I}}$, we find that $R_1$ rises to $4$ from $A_1$ as $p$ increases from $0$ to $1$.
This suggests that a larger $R_1$ is a signature of the higher $T_c$ caused by the singularity.
A similar tendency can be seen in the high-$T_c$ superconductivity in strongly correlated electron systems \cite{markiewicz_survey_1997}.
In $\mathcal{R}_{\text{III}}$, we find that $R_1$ decreases from $A_1$ to $0$ as $p$ decreases from $0$ to $-2$ and that $R_1$ is negligible for $p\leq-2$ as mentioned above.
We obtain $R_1=4\ln2$ for the honeycomb lattice $p=-1$, which is the same in Refs. \cite{uchoa_nodal_2005,kopnin_bcs_2008}.
Interestingly, at $p=0$ with a logarithmic singulariy, $R_1$ becomes $A_1$, which is a conventional value obtained for the non-singular DOSs.
In addition, three curves for the divergent, semi-metallic, and normal regions merge at point $p=0$ corresponding to the logarithimic divergent region.
This notable observation sheds light on the marginal properties of the logarithimic singularity.
The same results can be found for the other two universal ratios discussed below.

\subsection{$R_2$: specific heat jump at $T_c$}
\label{subsec_specificheatjump}

We next calculate $R_2$ defined by $\Delta C/S_N(T_c)$.
The finite value of $\Delta C$, the jump of the specifiec heat at $T_c$, is a clear evidence of the second order transitions because $C=-T\partial^2 F/\partial^2 T$.
On the other hand, entropy $S=-\partial F/\partial T$ is continuous at $T_c$, and thus $S_N(T_c)=S_S(T_c)[= S(T_c)]$.

Assuming $T_c/W \ll 1$, we obtain (see Appendix \ref{appendix_specficheat})
\begin{eqnarray}
\label{eq_R2}
R_2
&\equiv&
\left.
\frac{\Delta C(T)}{S_N(T)}
\right|_{T= T_c-0}
\nonumber
\\
&=&
\frac
{\left(
 \rho_{F} W+ a p B_{T_c} \left(\frac{T_c}{W}\right)^{- p}
 \right)^2}
{\left(
 B_0 \rho_{F}W + a B_1 \left(\frac{T_c}{W}\right)^{- p}
 \right)
 \left(
 \frac{2\pi^2}{3}\rho_{F}W
+
aB_2
\left(\frac{T_c}{W}\right)^{- p}
 \right)}
\nonumber
\\
&=&
\begin{cases}
 \frac{p^2 B_{T_c}^2}{B_1 B_2} (= \frac{pB_{T_c}}{B_2}R_0^2)
&(\mathcal{R}_{\text{I}} \cup \mathcal{R}_{\text{II}} \cup \mathcal{R}_{\text{III}}^>) \\
0& (\mathcal{R}_{\text{III}}^<) \\
\frac{12}{7\zeta(3)} (= \frac{3}{2\pi^2}A_0^2)
& (\mathcal{R}_{\text{IV}}). \\
\end{cases}
\end{eqnarray}
In the singular-DOS structure regions ($\mathcal{R}_{\text{I}} \cup \mathcal{R}_{\text{II}} \cup \mathcal{R}_{\text{III}}^>$), $R_2$ shows only $p$ dependence, whereas $R_2$ equals the conventional value in the normal region ($\mathcal{R}_{\text{IV}}$).
In the depleted DOS region ($\mathcal{R}_{\text{III}}^<$), $R_2$ is negligible because of $T_c/W \ll 1$ (see Appendix \ref{appendix_depletionDOSregion}).
These results are the same as those for universal ratio $R_1$.

Figure\ \ref{fig_universalratio}(b) shows the $p$ dependence of $R_2$.
For the divergent-structure region, $R_2$ is larger than the conventional value $12/7\zeta(3)(=1.43)$, whereas it is smaller for semi-metals.
We again find that the logarithmic divergent region locates in the marginal point among the other three regions---the divergent, semi-metallic, normal DOS regions.
The universal ratio $R_2$ increases as $p$ increases, and the maximaum is $3/2\ln2(=2.16)$ at $p=1$.
Equation \eqref{eq_entropy} suggests that entropy $S$ with a given $T$ increases as $p$ increases.
These features suggest that the jump of the specific heat $\Delta C$ increases with increasing $p$ and takes a maximum at $p=1$.
Namely, a large entropy enhanced by the singularity is drastically released at $T_c$.

The original definition of $R_2$ is given by $\Delta C/\gamma T_c$ \cite{bardeen_theory_1957}, and thus sometimes $R_2'\equiv \Delta C/C_N(T_c)$ is used for its generalized definition \cite{uchoa_nodal_2005}.
We obtain the general form of $R_2'$ as  $\frac{p^2 B_{T_c}^2}{(1-p)B_1 B_2} $ for the singular-structure region and $12/7\zeta(3)$ for the normal region.
This yields $8(\ln2)^2/9\zeta(3)$ for the honeycomb lattice $p=a=-1$, which is the same as that obtained in Ref. \cite{uchoa_nodal_2005}.
However, Fig.\ \ref{fig_universalratio}(b) shows that this $R_2'$ exhibits divergence at $p=1$ as $1/(1-p)$.
Such divergent behavior with a DFS has been pointed out in the theoretical study of the strained graphene \cite{uchoa_superconducting_2013}.
This divergence is in fact suppressed and is due to the approximation of $T_c/W \ll 1$.
More precisely, $R_2'$ is given by $(9a/2b\pi^2)(W/T_c)$ for $p=1$.
This suggests that the definition of $R_2'$ is no longer universal for the singular regions, which may be attributed to the discontinuity of the specific heat $C_N(T_c)\neq C_S(T_c)$ in contrast to $S_N(T_c)=S_S(T_c)$.

\subsection{$R_3$: critical magnetic field at $T=0$}
\label{subsec_criticalmagneticfield}

We next provide $R_3$ defined by $\frac{H_c(0)^2}{2|F_N(T_c)|}$.
The critical magnetic field at zero temperature $H_c(0)$ is related to the difference in the internal energy $\Delta E=E_S(T)-E_N(T)$ between the superconductor and normal conductor.
$\Delta E$ describes the energy gain from the formations of Cooper pairs.
Note that $R_3$ is defined with free energy, which is continuous at $T_c$ as $F_N(T_c)=F_S(T_c)$.
We obtain (see Appendix \ref{appendix_criticalmagneticfield})
\begin{eqnarray}
R_3
&\equiv&
\frac{H_c(0)^2}{2|F_N(T_c)|} 
\nonumber
\\
&=&
2\pi
\frac{\frac{1}{2}\rho_{F} W+a\frac{p}{2-p}B_{\Delta_0} (\frac{\Delta_0}{W})^{-p}}
     {2\left( \frac{\pi^2}{3}\rho_{F} W+ a \frac{B_2}{(2-p)} (\frac{T_c}{W})^{- p} \right)}
\left(\frac{2\Delta_0}{T_c}\right)^2
\nonumber
\\
&=&
\begin{cases}
\frac{4\pi pB_{T_c}}{B_2}\left(\frac{B_{\Delta_0}}{B_{T_c}}\right)^{\frac{2}{p}}
(=\pi \frac{pB_{T_c}}{B_2}R_1^2)
&(\mathcal{R}_{\text{I}} \cup \mathcal{R}_{\text{II}} \cup \mathcal{R}_{\text{III}}^>) \\
0& (\mathcal{R}_{\text{III}}^<) \\
6\pi e^{-2\gamma_E}(=\frac{3}{2\pi} A_1^2 )
& (\mathcal{R}_{\text{IV}}). \\
\end{cases}
\end{eqnarray}
In $\mathcal{R}_{\text{III}}^<$, we again find $R_3 \simeq 0$ for $T_c/W \ll1$ and $\Delta_0/W \ll1$ (see Appendix \ref{appendix_depletionDOSregion}).

Figure \ref{fig_universalratio}(c) shows $R_3$ as a function of $p$.
In the same manner as above, the magnitude of free energy $|F|$ increases as $p$ increases with fixed $T$.
Thus, the critical magnetic field $H_c$ becomes larger for larger $p$.
This behavior is consistent with the fact that the transition temperature is higher for larger $p$.

\subsection{Relations among universal ratios}
\label{subsec_relationbetweenuniversalratios}

We here exhibit a relation among universal ratios:
\begin{eqnarray}
\label{eq_relationbetweenURs}
\left(
 \frac{R_0}{R_1}
\right)^2
&=&
\pi\frac{R_2}{R_3}
\quad\quad
(\mathcal{R}_{\text{I}} \cup \mathcal{R}_{\text{II}} \cup \mathcal{R}_{\text{III}}^> \cup \mathcal{R}_{\text{IV}}).
\end{eqnarray}
We stress that this relation holds for all parameter regions except for ${\cal R}_{\text{III}}^<$ regardless of the presence or absence of the singular DOS structures.
Further, we argue that this equation provides a certification of consistency of our definition of universal ratios $R_2$ and $R_3$.

In ${\cal R}_{\rm III}^<$, we also find the relation
\begin{eqnarray}
\label{eq_relationbetweenURs2}
\left(
 \frac{R_0}{R_1}
\right)^2
&=&
\pi\frac{|p|+2}{2|p|}\frac{R_2}{R_3} \quad\quad ({\cal R}_{\text{III}}^<).
\end{eqnarray}
At $p\to -2$, this reproduces the relation shown above, while at $p\to -\infty$, the coefficient on the r.h.s. becomes one-half of the above.
In this depleted DOS region, all $R_n$ are negligible but finite, because $R_n \propto (T_c/W)^{-p/2-1} \ll 1$ or $\propto (T_c/W)^{-p-2} \ll 1$ for $p<-2$ 
\footnote{For $p=-2$, $R_n$ are proportional to $1/\ln(W/T_c)$ or $1/\sqrt{\ln(W/T_c)}$
}
(see Appendix \ref{appendix_depletionDOSregion}).
The above relation suggests that these power-law-$(T_c/W)$ dependences cancel each other out.

These relationships indicate that, regardless of the detailed structures of the DOS, the physical quantities at zero temperature and around $T_c$ should be related with each other.
This can be considered as the universal properties of the BCS type phenomena governed by the Fermi distribution.
Note that the depletion of the DOS slightly modifies the relation, but the nature of the transition is unchanged.

\section{Effective delta-functional singularity}
\label{sec_effective_flatbandsingularity}

Until now, we have provided analytical descriptions of the BCS theory for lattice fermions.
We have revealed that the DFS induces the highest transition temperature in the weakly interacting region, $T_c\propto aU$, which is the upper limit of $T_c$ instead of the conventional exponential form.
In actual materials, it seems to be hard to make the DFS at the Fermi energy.
On the other hand, the numerical simulations beyond the BCS theory provide a plausible mechanism of the emergence of this singularity.
We here demonstrate that correlation effects can induce the effective DFS in multi-energy systems such as multi-band systems with different bandwidths.
To this end, we calculate $\Phi(=\Delta/U)$ at zero temperature on a bipartite system.
This result obtained at $T=0$ allows us to infer the same behavior of transition temperature $T_c$ as discussed above \cite{noda_magnetism_2015}.

\subsection{DOS and Method}

We again consider the model described by the simple Hubbard Hamiltonian in Eq.\ (\ref{eq_hubbard}).
To emulate the multi-energy system within this model, we use a DOS consisting of two elliptic DOSs with different bandwidths.
\begin{eqnarray}
\label{eq_doublebethedos}
\rho(\omega)=a \rho^{\rm eli}(\omega,W_{\rm nar})+b\rho^{\rm eli}(\omega,W),
\\
\rho^{\rm eli}(\omega,W)=\frac{4}{\pi W}\sqrt{1-\left(\frac{W}{2\omega}\right)^2},
\end{eqnarray}
where $W_{\rm nar}=R_W W$ is the narrower bandwidth with the ratio of the different bandwidths $R_W(\leq 1)$ and $a (=1-b)$ denotes the ratio of the weights of bands (corresponding to the singularity weight).
The validity of this simplification (the use of the simplified DOS for the multiband systems) is discussed in our previous report \cite{noda_magnetism_2015}.

Figure\ \ref{fig_effectiveFBS}(a) shows the DOSs for $R_W=0.01, 0.1, 0.5$ and $1.0$  with $a$ of $0.5$.
A single elliptic structure appears for $R_W=1$, which is identical to the DOS of the Bethe lattice in infinite dimensions.
For $R_W< 1$, an additional narrower DOS structure appears with a width of $R_WW$.
When $R_W\to0$, the DOS of the narrow band becomes a delta-functional structure [see the case of $R_W=0.01$ in Fig. \ref{fig_effectiveFBS}(a)].
For a small but finite $R_W$, we naively expect that, when $U$ is larger than the narrower bandwidth, the narrow (but finite width) structure in the DOS can be regarded as delta functional.
We call this an effective DFS, which leads to a high-$T_c$.

The numerical method we employ here is based on the dynamical mean-field theory (DMFT)\ \cite{georges_dynamical_1996}.
In this theoretical framework, we solve the self-consistent equation that relates the original lattice model to an effective impurity model \cite{georges_dynamical_1996}, which allows us to precisely deal with local correlation effects.
We use the numerical renormalization group (NRG)\ \cite{bulla_numerical_2008} to solve the effective impurity model.
This method (DMFT plus NRG) has succeeded in elucidating the superconductor and magnetic transitions of correlated electrons \cite{bulla_numerical_2008}.

\subsection{Numerical results}

\begin{figure}[tb]
\includegraphics[clip,width=0.7\linewidth]{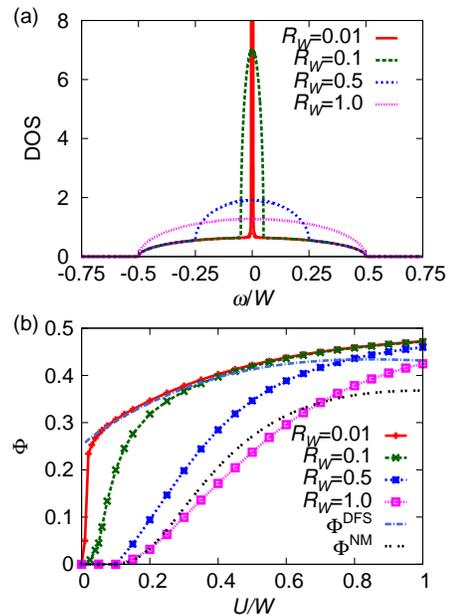}
\caption{(Color online) (a) DOS for $R_W=0.01, 0.1, 0.5$, and $1.0$ with $a=0.5$.  (b) SC order parameter $\Phi$ vs. $U$ for the same parameters.
The (dark-blue) dash-dotted line represents $\Phi^{\rm DFS}=aW\left/2bU \mathcal{W}_0\left(\frac{a e^{{W}/{bU}}}{2b}\right)\right.$,
and the (black) double-dashed line is $\Phi^{\rm NM}=(W/U)e^{-W/U}$.
}
\label{fig_effectiveFBS}
\end{figure}

Figure \ref{fig_effectiveFBS}(b) shows $\Phi$ at $T=0$ as a function of $U/W$ for $R_W=0.01, 0.1, 0.5, 1.0$ with $a$ of $0.5$.
We focus on the weakly interacting region. 
For $R_W=1.0$, our DMFT results show the conventional exponential behavior, which is consistent with the analytic results: $\Phi^{\rm NM}=(W/U) e^{-W/U}$.
As $R_W$ decreases, $\Phi$ in a small $U$ region becomes steep.
For $R_W=0.1$ and $0.01$, $\Phi$ exponentially increases for a small $U/W$ region in $U/W \leq 0.1$ and $\leq 0.01$, respectively.
This steep increase in $\Phi$ can be described as $\Phi^{\rm exp}=(W_{\rm nar}/U) e^{-W_{\rm nar}/U}$.
As $U/W$ further increases, the behavior of $\Phi$ for both $R_W=0.1$ and $0.01$ is similar, which can be described by the form obtained from Eq. \eqref{eq_Delta} with $p=1$:
\begin{equation}
\label{eq_mfbs}
\Phi^{\rm DFS}=\frac{aW}{2bU  \mathcal{W}_0\left(\frac{a e^{\frac{W}{bU}}}{2b}\right)} \simeq \frac{a}{2}+\frac{abU}{2W}\ln\left(\frac{2W}{aU}\right),
\end{equation}
which is shown as a dash-dotted line in Fig. \ref{fig_effectiveFBS}(b).

First, the order parameter exponentially develops as $\Phi^{\rm exp}$ for $U<W_{\rm nar}$, and then for $W_{\rm nar}<U<W$, it is described by $\Phi^{\rm DFS}$ with the Lambert W function, suggesting that the DFS is effectively induced by the interaction with the middle strengths of $W_{\text{nar}}<U<W$.
This is the effective DFS as we expected.
We note that there are two gap energy scales: fast developing $\Delta_{\text{exp}}(=\Phi_{\text{exp}}U)$ and slowly developing $\Delta_{\text{DFS}}(=\Phi_{\text{DFS}}U)$.
These results at $T=0$ indicate that $T_c$ as a function of $U$ shows the same linear $U$ dependence due to the effective DFS for $W_{\rm nar}<U<W$.
We thus argue that this effective DFS mechanism in multi-energy (multi-gap) systems provides a guide for designing high-$T_c$ SC materials.

We should note that the effective DFS can be found in the models for describing the actual systems, although we here employ the simplified DOS in Eq.\ (\ref{eq_doublebethedos}) for emulating a multi-energy system.
For example, in a three-dimensional layered Lieb lattice, which is realized in cold atom experiments \cite{taie_matter-wave_2015}, the sublattice structure induces multi-energy scales, and the layered structure enables us to control the band structure \cite{noda_magnetism_2015}.
Our previous numerical calculations correctly dealing with the sublattice structure show a good agreement with the analytical description, suggesting that our modeling of the DOS is valid for a sublattice system \cite{noda_magnetism_2015}.
Another example of the mechanism has been pointed out in Ref.\ \cite{kuroki_high-$t_c$_2005}: The authors have investigated a high-$T_c$ SC mechanism on the Hubbard ladder system with a repulsive interaction with numerical calculations based on the fluctuation exchange approximation.
Although model and calculation details are different, the main statement of this study is consistenent with what we pointed out: a correlated electron system with a coexistence of narrow and wide bands can induce higher $T_c$.

\section{Summary}
\label{sec_summary}

We investigated superconducting transitions in lattice fermions based on the BCS theory with a modeling of the DOS, which describes divergent and semi-metallic structures.
We derived an analytical solution of $T_c$ and $\Delta$ of all systems described by this DOS at half filling.
This simple modeling captures the competition between polynomial and exponential responses of the degenerate fermions around the Fermi energy, which is crucial in phase transitions in lattice fermions.

We revealed that the asymptotic forms of $T_c$ with $U$ are classified into four types.
In divergent and logarithmic-divergent regions ${\cal R}_{\text{I}}\cup{\cal R}_{\text{II}}$, these functional forms of $T_c$ include singularity weight $a$ as $aU$.
This suggests that a large $a$ induces higher $T_c$.
On the other hand, in ${\cal R}_{\text{III}}$, the semi-metallic structures induce quantum phase transitions with finite interactions.
In normal region ${\cal R}_{\text{IV}}$, we found that the sign of singularity weight $a$ determines whether the interaction is enhanced or suppressed.
These various asymptotic forms originate from the Fermi degeneracy.

We provided generalized definitions of universal ratios in the BCS theory to deal with the singular structures of the DOS.
These ratios no longer show the single values but can be parametrized by a single variable that characterizes the singular DOS structures.
We uncovered relationship among these ratios, which assures that the superconducting transition is of the second order with divergent and semi-metallic DOS structures.

We numerically elucidated that in a two-band system correlation effects can incorporate the two (gap) energy scales, which leads to the effective delta-functional singularity for $W_{\text{nar}}<U<W_{\text{wide}}$.
This effective singularity enhances $T_c$, which can be described as $T_c\propto U$.
This is the upper limit elucidated by our description.
This phenomenon generally appears in a multi-energy system and may be used as a guideline for material-designs for obtaining the high-$T_c$ SC materials.


\begin{acknowledgments}
This work was partly supported by JSPS KAKENHI (Grant No. 25287104).

\end{acknowledgments}


\appendix


\section{Lambert W function $\mathcal{W}_n(x)$}
\label{appendix_Lambert}

The Lambert W function $\mathcal{W}_n(x)$ is the inverse function of $xe^x$, and its definition is written as $x=\mathcal{W}_n(x)e^{\mathcal{W}_n(x)}$ \cite{corless_lambertw_1996}.
This function can be regarded as a generalization of the logarithmic function defined by $x=e^{\ln(x)}$, which is the inverse function of $e^x$.
With a restriction of real-valued $\mathcal{W}_n(x)$ and $x$, the following two branches are defined as shown in Fig.\ \ref{fig_lambert}: the principal branch $\mathcal{W}_0(x)$ defined for $[-1/e, \infty)$ and the negative branch $\mathcal{W}_{-1}(x)$ for $[-1/e,0)$.
Note that the Lambert W function is multivalued for $[-1/e,0)$.

As shown in Fig.\ \ref{fig_lambert}, the asymptotic expansions of these branches are given by
\begin{subnumcases}
{\mathcal{W}_0(x)=}
\label{eq_lambert0_around0}
x \hspace{2cm} (x \simeq 0) \\
\label{eq_lambert0_aroundinfty}
\ln x-\ln(\ln x) \hspace{0.2cm} (x\to\infty) \\
\label{eq_lambert0_aroun1/e}
-1+\sqrt{2e\left(x+\frac{1}{e}\right)} \nonumber \\
\hspace{2cm} (x\simeq -1/e),
\end{subnumcases}
and
\begin{subnumcases}
{\mathcal{W}_{-1}(x)=}
\label{eq_lambertm1_around1/e}
-1-\sqrt{2e\left(x+\frac{1}{e}\right)} \\
\hspace{2cm}(x\simeq -1/e)
\nonumber
\\
\label{eq_lambertm1_around0}
\ln(-x)-\ln(-\ln(-x)) \\
\hspace{2cm}(x\to0).
\nonumber
\end{subnumcases}
Interestingly, these expansions lead to four different asymptotic $U$-dependences of $T_c$ and $\Delta$ (see Sec.\ \ref{subsec_asymptonic} and Appendix \ref{appendix_asymptoticp}).

The Lambert function can describe both polynomial and exponential behaviors, which naturally appear in physical systems.
Actually, in the past, P. Nozi\'eres and F. Pistolesia have employed this function to deal with semiconductor-superconductor transitions without the name ``Lambert W function" \cite{nozieres_semiconductors_1999}.
These facts indicate that the Lambert function could be commonly used for a description of phase transitions of lattice fermions.

\section{Integral for $T_c$}
\label{appendix_DerivationTanhIntegral}

To derive Eq.\ (\ref{eq_gap_Tc}), we use the following integral equation:
\begin{multline}
\label{eq_int_tanh}
I(p)=\int_{0}^{y}\omega^{-p}\frac{\tanh(\omega)}{\omega}d\omega
\\
\sim
-\frac{1}{p}y^{-p}+(4^{p+1}-2^{p+1})\Gamma(-p)\zeta(-p)
\\
{\rm for}\ \  y\gg 1 \text{ and } p<1,
\end{multline}
where $\Gamma(x)$ and $\zeta(x)$ represent the gamma and Riemann zeta functions, respectively.
By applying coefficient $a[(1-p)/p](W/4T)^p$, and by substituting $W/4T$ into $y$, we can determine the parameters in Eq.\ (\ref{eq_gap_Tc}). Namely, $\chi$ and $B_{T_c}$ can be obtained from the first and second terms in the second line of Eq.\ \eqref{eq_int_tanh}, respectively.
For $p\to 1$, $\Gamma(-p)$ diverges as $1/(1-p)$, which sets the bound of $p<1$.
This divergence is canceled by the coefficient $1-p$, and thus $B_{T_c}$ is defined for $p\leq 1$.

We briefly comment about the $p\to 0$ limit of Eq.\ (\ref{eq_int_tanh}).
For this limit, Eq.\ (\ref{eq_int_tanh}) yields the well-known integral $I(0)=\ln(4e^{\gamma}y/\pi)$, which leads to the third term on the r.h.s. of Eq.\ (\ref{eq_gap_Tc}).
This can be obtained by considering that the second term in (\ref{eq_int_tanh}) reduces to $\ln(4e^{\gamma}/\pi)+1/p$ for $p\to 0$ and that $\lim_{p\to0}(1-y^{-p})/p=\ln y$.
On the other hand, the singular DOS in Eq.\ (\ref{eq_generalizeddos_singular}) leads to a logarithmic DOS for $p=0$.
We can correctly obtain the integral equation $\int d\omega \log(\omega) \tanh(\omega)/\omega$ from $\lim_{p\to0} [(1-p)/p][I(p)-I(0)]$.

To make this paper self-contained, we briefly explain the derivation of the above integral (\ref{eq_int_tanh}).
We divide the integral region as follows:
\begin{eqnarray}
I(p)
&=&
\underbrace{\int_0^1 dx \frac{ \tanh x}{x^{p+1}} }_{=I_1}
+
\int_1^{y} dx \frac{1}{x^{p+1}}
-
\underbrace{\int_1^{y} dx \left(\frac{2x^{-p-1} }{e^{2x}+1} \right)}_{=I_2}
\nonumber
\\
&=&
-\frac{1}{p}(y^{-p}-1)+I_1-I_2,
\label{eq_inttanhI0}
\end{eqnarray}
where we use $\tanh(x)=1-2/(e^{2x}+1)$.
In the following, we approximately take $y\to\infty$ in integral $I_2$. 
Error in this approximation decreases exponentially for a large $y$.

Integral $I_1$ is written as
\begin{eqnarray}
I_1 \nonumber
&=&
2\int_0^1 dx \frac{\sinh(x)}{x^{p+1}}\sum_{n=0}^{\infty}(-1)^ne^{-(2n+1)x}
\\
&=&
(4^{p+1}-2^{p+1})\Gamma(-p)\zeta(-p)
\nonumber
\\
&&
+
\frac{4}{p}\sum_{n=1}^{\infty}(-1)^nnE_{p}(2n)
+
\frac{2}{p(1+e^2)}
-
\frac{1}{p},
\label{eq_inttanhI1}
\end{eqnarray}
where $E_{p}(x)$ is the exponential integral defined as $E_{p}(x)=\int_1^{\infty}dt\frac{e^{-xt}}{t^p}$.
Here, we use the following equations to obtain the final result of Eq.\ \eqref{eq_inttanhI1}: $2\int_0^1 dx \frac{\sinh(x)}{x^{p+1}}e^{-(2n+1)x}
=2^p\Gamma(-p)[n^p-(n+1)^p]-E_{p+1}(2n)+E_{p+1}(2n+2)$,
$E_{p+1}(2n)=\frac{e^{-2n}}{p}-\frac{2n}{p}E_p(2n)$, and $\sum_{n=1}^{\infty}(-1)^n n^p=(2^{1+p}-1)\zeta(-p)$.

Integral $I_2$ is calculated by partial integration as
\begin{eqnarray}
I_2
&=&
\frac{2}{p(1+e^2)}
-
\frac{4}{p}\int_1^{\infty} dx \frac{1}{x^p}\frac{e^{-2x}}{(1+e^{-2x})^2}
\nonumber
\\
&=&
\frac{2}{p(1+e^2)}
+
\frac{4}{p}\sum_{n=1}^{\infty}(-1)^{n}n E_p(2n),
\label{eq_inttanhI2}
\end{eqnarray}
where we use $\frac{r}{(1+r)^2}=\sum_{n=1}^{\infty}(-1)^{n+1}nr^n$.
Substituting Eqs. (\ref{eq_inttanhI1}) and (\ref{eq_inttanhI2}) into Eq. (\ref{eq_inttanhI0}), we obtain Eq. (\ref{eq_int_tanh}).

\section{Integral for $\Delta_0$}
\label{appendix_int}

We use the following integral equation to derive Eq. (\ref{eq_gap_delta}):
\begin{eqnarray}
\label{eq_int_hypergeom_rho}
&&
\int_0^{W/2}d\omega
\omega^{-p}
\frac{1}{\sqrt{\omega^2+\Delta_0^2}}
\nonumber
\\
&=&
\left(\frac{2}{W}\right)^p
\frac{W}{\Delta_0}
\frac{_2F_1\left(\frac{1}{2},\frac{1- p}{2};\frac{3- p}{2};-\frac{W^2}{4 \Delta_0 ^2}\right)}{2(1-p)}
\quad
(p<1)
\nonumber
\\
&=&
\left(\frac{2}{W}\right)^p
\left\{
-
\frac{1}{p}
\right.
+\frac{\Gamma \left(\frac{1-p}{2}\right)
       \Gamma \left(\frac{p}{2}\right)}
      {2^{p+1}\sqrt{\pi }}
\left(\frac{\Delta_0}{W}\right) ^{- p}
\nonumber
\\
&& \,\,\,
\left.
+\frac{2}{2+p}\left(\frac{\Delta_0}{W}\right)^2
+O\left(\frac{\Delta_0}{W}\right)^4
\right\},
\end{eqnarray}
where $_2F_1(\alpha,\beta;\gamma;x)$ denotes the hypergeometric function.
In the same way as above (Appendix \ref{appendix_DerivationTanhIntegral}), by applying coefficient $a[(1-p)/p](W/2)^p$, we obtain the first and the second terms on the r.h.s. of Eq.\ (\ref{eq_gap_delta}).

Equation\ (\ref{eq_int_hypergeom_rho}) shows that the second leading term changes at $p=-2$ with decreasing $p$ from $\propto (\Delta_0/W)^{-p}$ to $\propto (\Delta_0/W)^{2}$, and then it remains $\frac{2}{2+p}(\Delta/W)^2$ for $p<-2$.
As a result, $\bar{p}$ and $B_{\Delta_0}$ show complex $p$ dependences as denoted in Eqs. \eqref{eq_Delta_parameter_p} and \eqref{eq_Delta_parameter_BDelta0}.
For $p=0$, Eq.\ (\ref{eq_int_hypergeom_rho}) reduces to $\ln(W/\Delta_0)+(\Delta_0/W)^2$, the first term of which yields the third term on the r.h.s. of Eq.\ (\ref{eq_gap_delta}).
Note that the second term $(\Delta_0/W)^2$ is included in $B_{\Delta_0}$ for $p\leq -2$.

\section{Asymptotic expansions}
\label{appendix_asymptoticp}

We here explain how the asymptotic forms of $T_c$ for the weakly interacting limit $U/W \ll 1$ are determined from the unified expression of $T_c$ with the Lambert W function in Eq. (\ref{eq_Tc_2}). 
We consider the argument of $W_n(X)$ in Eq. (\ref{eq_Tc_2}) given by $X=\frac{apB_{T_c}}{W \rho_F} A_{T_c}^{p} e^{p/\rho_F \bar{U}}$.
As $U$ decreases down to the weak interaction limit, $U/W\to 0$, the argument $X$ goes to either $\pm0$, $-1/e$, or $\infty$ depending on $p$ and $a$.
The asymptotic behavior of ${\cal W}_n(X)$ around these points shown in Eqs. (A1) and (A2) determine the forms of $T_c$ for $U/W\ll 1$ in Eq. (\ref{eq_asymptonic}).
Considering the $(a,p)$-dependences of $X$, we define the four regions ${\cal R}_{n} (n=$I, II, III, and IV) in Table \ref{table_diagram}.
For negative $X \le 0$, the Lambert function is multivalued, and then we have to carefully choose either ${\cal W}_0(X)$ or ${\cal W}_{-1}(X)$ to obtain a correct form of $T_c$.
Note that the wrong choice leads to the unphysical solution of the gap equation.
In the following, we provide the detailed derivations of the asymptotic behavior in each region ${\cal R}_n$, which can be straightforwardly extended to the asymptotic forms of $\Delta_0$.

In divergent-structure region $\mathcal{R}_{\text{I}}$, $X$ goes to $\infty$ for $p\leq a$, while it goes to $-0$ for $p>a$, which are dominated by $e^{p/\rho_F\bar{U}}$, where $p>0$, $B_{T_c}>0$, and $\bar{U} \simeq U(>0)$.
For $p\leq a$, we employ the expansion $\mathcal{W}_{0}(X \sim \infty) \simeq (p/\rho_F\bar{U})+\ln (a B_{T_c}\bar{U}/W)+p\ln A_{T_c}$.
Substituting this into Eq.\ (\ref{eq_Tc_2}), we obtain $T_c/W =(aB_{T_c}\bar{U}/W)^{1/p}$.
For $p>a$, since $X \leq 0$, we should choose either $\mathcal{W}_{0}(X \sim -0) \simeq X (\sim 0)$ or $\mathcal{W}_{-1}(X \sim -0) \simeq (p/\rho_F\bar{U})+\ln (a B_{T_c}\bar{U}/W)+p\ln A_{T_c}$.
The former yields the unphysical solution $T_c/W = (1/ A_{T_c})e^{-1/\rho_F\bar{U}} \to \infty$ because of a negative $\rho_F$, whereas the latter yields the physical $T_c$ in the same form as for $p\leq a$.
Note that the expansions ${\cal W}_0(x \simeq \infty)$ and ${\cal W}_{-1}(x\simeq 0)$ are written as the single from $\mathcal{W}_{n}(x)=\ln((-1)^nx)-\ln((-1)^n\ln((-1)^nx))$.
This form provides the physical solution of the power-law-$U$ dependences of $T_c$ in ${\cal R}_{\text{I}}$: $T_c/W=(aB_{T_c}{U}/W)^{1/p}$.

In log-divergent structure region $\mathcal{R}_{\text{II}}$, $X$ goes to $-1/e$, where $\rho_F$, $\chi$, and $B_{T_c}$ diverge for $p\to 0$, but they totally cancel out each other.
The expansion of $X$ around $p=0$ is given by $-1/e+(p^2/a^2e)(aW/U+b^2/2+a^2A_{\text{log}}/2)+O(p^3)$, where $b=1-a$ and $A_{\text{log}}=\gamma^2-\pi^2/4+2\ln^2 2+2\gamma_{1}$ with the Stieltjes constant $\gamma_1(=-0.0728)$.
Then, we obtain the multivalued asymptotic forms ${\cal W}_{n}(X \sim -1/e)\simeq -1 +(-1)^n (p/a)\sqrt{2aW/U+b^2+a^2A_{\text{log}}}$.
We also need the expansion of $-1/\rho_F \bar{U}$ around $p=0$, which is given by $1/p+b/a+O(p)$.
Substituting these expansions into Eq.\ (\ref{eq_Tc_2}), we obtain $T_c$ for the log-divergent structure region: $T_c/W=\exp\left(\frac{b}{a}+(-1)^n\frac{b}{a}\sqrt{1+\frac{2aW}{b^2 U}+\frac{a^2 A_{\text{log}}}{b^2}}\right)\big/ A_{T_c}$.
The principal branch yields the unphysical $T_c (\to \infty)$, while
the negative branch provides the physical $T_c\simeq (1/ A_{T_c})e^{-\sqrt{2W/aU}}$.

In semi-metallic structure region $\mathcal{R}_{\text{III}}$, $X$ goes to $\infty$ because $\rho_F\to +0$, $p=a<0$, and $B_{T_c}>0$.
We obtain $T_c/W =(pB_{T_c}\bar{U}/W)^{1/p}$, which is the same as the form in $\mathcal{R}_{\text{I}}$.
For a small $U\simeq 0$, $\bar{U} (\simeq U)$ is positive, and thus $pB_{T_c}\bar{U}$ is negative, indicating that $T_c/W$ is complex and thus an unphysical solution.
For a larger $U \ge -1/a\chi$, the sign of $\bar{U}$ becomes negative.
Thus, the above form yields the physical solution.
This clarifies that the SC phase transition occurs at the critical point $U_c\equiv -1/a\chi [=|p|/(|p|+1)W]$.
It is convenient to rewrite the form of the physical solution as $T_c/W=\big((U-U_c)W/|p|B_{T_c}U_c^2\big)^{1/|p|}$.

In normal structure region $\mathcal{R}_{\text{IV}}$, $X$ goes to $\pm0$ for a negative (positive) $a$ because $e^{p/\rho_F\bar{U}}\to 0$, $p<0$, $B_{T_c}>0$,  and $\bar{U}\sim U(>0)$.
For $a>0$,  the negative branch yields an unphysical $T_c/W (\to \infty)$.
For both positive and negative $a$, the physical solution is obtained from the asymptotic form ${\cal W}_0(X\simeq 0)\simeq X (\simeq 0)$, which reproduces the well-known exponential form $T_c/W = (1/ A_{T_c}) e^{-1/\rho_F\bar{U}}$.

\section{Gap energy around $T_c$}
\label{appendix_DeltaNearTc}

The gap energy $\Delta$ around $T_c$ in Eq.\ (\ref{eq_DeltaaroundTc_line1}) in ${\cal R}_{\rm tot}$ except for ${\cal R}_{\rm III}^<$ is determined from the following gap equation:
\begin{eqnarray}
\label{eq_gaparoundTc_total_appendix}
&& \int_{-\infty}^\infty d\omega \rho(\omega)
\frac{1}{2\omega}\tanh\left(\frac{\omega}{2T}\right)
-  \int_{-\infty}^\infty d\omega \rho(\omega)
\frac{1}{2\omega}\tanh\left(\frac{\omega}{2T_c}\right) \nonumber
\\
&&=
\sum_{n=0}^\infty \int_{-\infty}^\infty d\omega\rho(\omega) \frac{2T\Delta^2}{(\omega^2+((2n+1)\pi
T)^2)^2},
\end{eqnarray}
where we use $\frac{\tanh(\sqrt{x^2+\Delta^2}/2T)}{2\sqrt{x^2+\Delta^2}}-\frac{\tanh(x/2T)}{2x}
\sim -\sum_{n=0}^{\infty}\frac{2T\Delta^2}{(x^2+((2n+1)\pi T)^2)^2}+{\cal O}(\Delta^4)$,
and $\int_{-\infty}^\infty d\omega \rho(\omega) \frac{1}{2\omega}\tanh(\frac{\omega}{2T_c})=1/U$.
Equation \eqref{eq_gaparoundTc_total_appendix} does not hold for $p\le-2$ with $a\not=0$ because the second line of Eq.\ \eqref{eq_gaparoundTc_total_appendix} diverges at $p=-2$ with $a=p$.

Here we mention the $p=0$ limit of $R_0$ in Eq. (\ref{eq_DeltaaroundTc}), where $R_0$ is given by $R_0\simeq\sqrt{ \frac{(1-a)+ a \ln(W/T_c)}{B_0 [(1-a) +a \ln(W/T_c)]} }$.
It can be further reduced to $\sqrt{1/B_0}$ for both $a=0$ ($\in {\cal R}_{\rm IV}$) and $a\not=0$ ($\in {\cal R}_{\rm II}$). Note that $\sqrt{p B_{T_c}/B_1}$ coincides with $\sqrt{1/B_0}$ for $p=0$. Thus, $R_0=\sqrt{p B_{T_c}/B_1}$ is defined for all of the regions with singular structures $\mathcal{R}_{\text{I}} \cup \mathcal{R}_{\text{II}} \cup \mathcal{R}^>_{\text{III}}$.

For $a=p\leq-2$ (in ${\cal R}_{\rm III}^<$), $\Delta$ around $T_c$ in Eq.\ (\ref{eq_DeltaaroundTc_line3}) is obtained from the following gap equation:
\begin{eqnarray}
\label{eq_gaparoundTc_depleted_appendix}
&& \int_{-\infty}^\infty d\omega \rho(\omega)
\frac{1}{2\omega}\tanh\left(\frac{\omega}{2T}\right)
-
\int_{-\infty}^\infty d\omega \rho(\omega)
\frac{1}{2\omega}\tanh\left(\frac{\omega}{2T_c}\right)
\nonumber
\\
&&
=
\frac{\partial}{\partial T} 
\int_0^{\infty}
d\omega \rho(\omega) 
\frac{-\Delta^2}{2\omega^3}T
\tanh\left(\frac{\omega}{2T}\right)
\end{eqnarray}
where we use an expansion around $\Delta=0$: $\frac{\tanh(\sqrt{x^2+\Delta^2}/2T)}{\sqrt{x^2+\Delta^2}}-\frac{\tanh(x/2T)}{x}\sim\frac{\partial}{\partial T}[-\frac{\Delta^2}{2\omega^3}T \tanh(x/2T)]+{\cal O}(\Delta^4)$.
Equation \eqref{eq_gaparoundTc_depleted_appendix} holds for $a=p<-1$ because of the bound of the integral in the second line in Eq.\ \eqref{eq_gaparoundTc_depleted_appendix}.
Note that both Eqs. \eqref{eq_gaparoundTc_total_appendix} and \eqref{eq_gaparoundTc_depleted_appendix} hold for $-2<p<-1$ with $p=a$ and yield the same result, $\sqrt{pB_{T_c}/B_{1}}$, in ${\mathcal R}_{\text{III}}^>$.

\section{Specific heat difference at $T_c$: $\Delta C$}
\label{appendix_specficheat}

The discrete jump of specific heat at the transition temperature $\Delta C$ characterizes the second universal ratio $R_2 [\equiv \Delta C/S(T_c)]$.
We calculate $\Delta C$ as
\begin{eqnarray}
&&\Delta C=[C_S(T)-C_N(T)]_{T=T_c-0}
\nonumber
\\
&&=
\int d\omega \rho(\omega)\frac{\partial f(\omega)}{\partial \omega}
\left.
\frac{\partial \Delta(T)^2}{\partial T}\right|_{T=T_c-0}
\nonumber
\\
&&=
\frac{T_c}{W}
\frac
{(\rho_{F}W + a p B_{T_c} (T_c/W)^{- p})^2}
{(B_0 \rho_{F}W + a B_1 (T_c/W)^{- p})},
\end{eqnarray}
which does not hold for $R_{\text{III}}^<$.
We use Eqs. (\ref{eq_int_tanh}) and (\ref{eq_DeltaaroundTc_line1}) to obtain $\Delta C$.

The finite $\Delta C$ reflects the fact that the SC transition is of the second order.
Namely, the second-order derivative of free energy, which is the specific heat $C=-T\partial^2 F/\partial^2 T$, is discontinuous at the transition point.
In contrast, the entropy, which is the denominator of $R_2$, is continuous at $T_c$: $S_S(T_c)=S_N(T_c)$.
If we use $C_N(T_c)$ as the denominator of the universal ratio, it diverges at $p=1$ as mentioned in the main text.

\section{Critical magnetic field at $T=0$: $H_c(0)$}
\label{appendix_criticalmagneticfield}

The critical magnetic field  $H_c(T)$ characterizes the third universal ratio $R_3=H_c(T=0)^2/2|F_N(T_c)|$.
This quantity is related to the difference in free energies between the superconducting and normal conducting states: $\frac{H_c(T)^2}{8\pi}=F_N(T)-F_S(T)$.
At zero temperature, the free energy reduces to the internal energy $E$, and thus we obtain 
\begin{eqnarray}
\frac{H_c(0)^2}{8\pi}
&=&\Delta E
=E_N(0)-E_S(0)
\nonumber
\\
&=& \int_{-W/2}^0 d\omega \rho(\omega)
\left(\frac{2\omega^2+\Delta_0^2}{\sqrt{\omega^2+\Delta_0^2}}+2\omega\right)
\nonumber
\\
&=&
\frac{1}{2} \rho_{F} \Delta_0^2
-a\frac{ \Gamma\left(\frac{3}{2}-\frac{ p}{2}\right) \Gamma\left(\frac{ p }{2}-1\right) }{2^{1+ p}\sqrt{\pi}}
\frac{\Delta_0^{2- p}}{W^{1- p}}
\nonumber
\\
&&
-\frac{2-3a+ p}{2(2+ p)}\frac{\Delta_0^4}{W^3}
+O\left(\frac{\Delta_0}{W}\right)^6,
\end{eqnarray}
where the second leading term changes at $p=-2$.
Such a special property at $p=-2$ results from the fact that $\partial \rho(\omega)/\partial \omega |_{\omega\to 0}$ is discontinuous for $-2<p$, and this behavior disappears for $p\leq-2$.
A similar anomaly is found in the gap equation at zero temperature for deriving $\Delta_0$.
Note that the steep Fermi surface with a finite jump at $T=0$ is relevant for this special property, and thus such a property cannot be found at finite temperatures with a smooth Fermi surface.

The energy difference $\Delta E$ can be characterized by $\Delta E= \Delta_0 N_{\rm pair}$, meaning that $N_{\rm pair}$ particles form pairs with binding energy $\Delta_0$.
For $p=1$, we obtain $N_{\rm pair}=-\frac{1}{2}\frac{\Delta_0}{W}(1-a)-\frac{a}{2}$.
$a/2$ corresponds to the number of particles forming pairs from the flat-band structure.

\section{Depleted DOS region ${\cal R}_{\rm III}^<$ }
\label{appendix_depletionDOSregion}

In ${\cal R}_{\rm III}^<$ ($a=p\leq-2$), the DOS is semi-metallic and smoothly decreases around $\omega=0$, suggesting that the DOS around $\omega=0$ is very dilute.
The depletion of DOSs causes the anomalous behavior in all $R_{n=0,1,2,3}$, which are given by:
\begin{eqnarray}
R_0&=& \left( \frac{-pB_{T_c}}{B_{\Delta_0}} \right)^{1/2} \left(\frac{T_c}{W}\right)^{-p/2-1}, \nonumber\\
R_1&=&  2\left(\frac{B_{T_c}}{B_{\Delta_0}} \right)^{1/2} \left(\frac{T_c}{W}\right)^{-p/2-1}, \nonumber \\
R_2&=& \frac{-p^2 B_{T_c}^2}{B_2B_{\Delta_0}} \left(\frac{T_c}{W}\right)^{-p-2}, \nonumber\\
R_3&=& \frac{2\pi(p-2)B_{T_c}^2}{B_2B_{\Delta_0}}\left(\frac{T_c}{W}\right)^{-p-2}, \nonumber
\end{eqnarray}
where $B_2<0$, $B_{T_c}>0$, and $B_{\Delta_0}=2(1-p)/p(p+2) >0$.
Using these equations, we obtain the modified universal relationship in Eq.\ \eqref{eq_relationbetweenURs2}.

\section{Table of thermodynamic quantities}
\label{appendix_thermodynamicqunatities}

For convenience, in Table\ \ref{table_thermodynamics}, we summarize $T_c, \Delta_0$, universal ratios, and thermodynamic quantities of specific DOSs: the delta-functional divergent DOS (flat-band systems, such as the Lieb lattice), the logarithmic divergent DOS (the square lattice), and the linearly vanishing DOS (the honeycomb lattice).
These examples are representatives of each classified regions shown in Table \ref{table_diagram}.

\onecolumngrid
\begin{center}
\begin{table}[hb]
  \begin{center}
    \caption{Thermodynamic properties of several systems. $A_{T_c}=\pi e^{-\gamma_E}$ and $A_{\text{log}}=\gamma_E^2-\pi^2/4+2\ln^2 2+2\gamma_{1}$, where $\gamma_E(=0.577)$ and $\gamma_1(=-0.0728)$ is the Euler's and Stieltjes's constant, respectively. $A_{\text G}(=1.28)$ represents the Glaisher's constant, which has a relation: $\ln A_{\text G}=1/12-\zeta'(-1)$. There is the normalized condition: $a+b=1$.}
    \begin{tabular}{cc|ccc} \hline
\label{table_thermodynamics}
       & \hspace{-0.5cm} non-singular lattices & Lieb lattice & square lattice & honeycomb lattice \\
       & $\mathcal{R}_{\text{IV}}$ & $ p= 1 (\in \mathcal{R}_{\text{I}})$ & $ p= 0 (\in \mathcal{R}_{\text{II}})$ & $ p=a=-1$ $(\in \mathcal{R}_{\text{III}})$ \\
      \hline\hline
      $T_c/W$ & $e^{-1/\rho_{F}\bar{U}}/A_{T_c}$ & $a\big/4b \mathcal{W}_0\left(aA_{T_c} e^{\frac{W}{bU}}/4b \right)$ & $\exp\left(\frac{b}{a}-\frac{b}{a}\sqrt{1+\frac{2aW}{b^2 U}+\frac{a^2 A_{\text{log}}}{b^2}}\right)\big/A_{T_c}$ & $(U-W/2)/4U\ln2 $ \\
      $T_c/W$ (asym.)&  & $aU/4W$ & $e^{-\sqrt{2W/aU}}/A_{T_c}$ &  \\
      $\Delta_0/W$ & $e^{-1/\rho_{F}\bar{U}}$ & $a\big/2b  \mathcal{W}_0\left(a e^{\frac{W}{bU}}/2b\right)
$ & $\exp\left(\frac{b}{a}-\frac{b}{a}\sqrt{1+\frac{2aW}{b^2 U}-\frac{a^2 \pi^2}{6b^2}}\right)$ & $(U-W/2)/2U$ \\
      $\Delta_0/W$(asym.) &   & $aU/2W$ & $e^{-\sqrt{2W/aU}}$&   \\
      \hline
      $R_0$  & $\sqrt{8\pi^2/7\zeta(3)}$ & $2\sqrt{3}$ & $\sqrt{8\pi^2/7\zeta(3)}$ & $2\sqrt{2\ln2}$ \\
      $R_1$  & $2\pi e^{-\gamma_E}$ & $4$ & $2\pi e^{-\gamma_E}$ & $4\ln2$ \\
      $R_2$  & $12/7\zeta(3)$ & $3/2\ln2$ & $12/7\zeta(3)$ & $16(\ln2)^2/9\zeta(3)$ \\
      $R_3$ & $6\pi e^{-2\gamma_E}$ & $2\pi/\ln2$ & $6\pi e^{-2\gamma_E}$ & $32\pi(\ln2)^3/9\zeta(3)$\\
      \hline
      $C_N(T)$ & $\frac{2\pi^2}{3}\rho_{F}T$ & $\frac{2\pi^2}{3}\frac{T}{W}(1-a)$ & \shortstack{$\frac{2\pi^2}{3}\frac{T}{W}\{1+a\ln(W/T)$ \\ $-a\ln(2^3 e^{5/2}\pi/A_{\text G}^{12})\}$ } & $72\zeta(3)\left(\frac{T}{W}\right)^2$ \\
      $S_N(T)$ & $\frac{2\pi^2}{3}\rho_{F}T$ & $\frac{2\pi^2}{3}\frac{T}{W}(1-a)+2a\ln2$ & \shortstack{$\frac{2\pi^2}{3}\frac{T}{W}\{1+a\ln(W/T)$ \\ $-a\ln(2^2 e^{3/2}\pi/A_{\text G}^{12})\}$ } & $36\zeta(3)\left(\frac{T}{W}\right)^2$ \\
      \hline
      $\frac{H_c^2(0)}{8\pi^2}(=\Delta E)$ & $\frac{\rho_{F}\Delta^2}{2}$ & $\frac{\Delta^2}{2W}(1-a)+\frac{a\Delta}{2}$ & $\frac{\Delta^2}{2W}[(1-2a)+a\ln(W/\Delta)]$ & $\frac{4\Delta^3}{3W^2}$ \\
      \hline
    \end{tabular}
  \end{center}
\end{table}
\end{center}
\twocolumngrid

%

\end{document}